# 3-D macro/microporous-nanofibrous bacterial cellulose scaffolds seeded with BMP-2 preconditioned mesenchymal stem cells exhibit remarkable potential for bone tissue engineering

Swati Dubey[a], Rutusmita Mishra[b], Partha Roy[b] and R. P. Singh[a]*

[a]Microbial Biotechnology Laboratory, Department of Biotechnology, Indian Institute of Technology Roorkee, Roorkee-247667, India.

[b]Molecular Endocrinology Laboratory, Department of Biotechnology, Indian Institute of Technology Roorkee, Roorkee-247667, India.

*Corresponding author. Tel.: +91 1332 285792; Fax: +91 1332 273560.

E-mail addresses: r.singh@bt.iitr.ac.in (R.P. Singh); sdubey1@bt.iitr.ac.in (Swati Dubey).

**Abstract**

Bone repair using BMP-2 is a promising therapeutic approach in clinical practices, however, high dosages required to be effective pose issues of cost and safety. The present study explores the potential of low dose BMP-2 treatment via tissue engineering approach, which amalgamates 3-D macro/microporous-nanofibrous bacterial cellulose (mNBC) scaffolds and low dose BMP-2 primed murine mesenchymal stem cells (C3H10T1/2 cells). Initial studies on cell-scaffold interaction using unprimed C3H10T1/2 cells confirmed that scaffolds provided a propitious environment for cell adhesion, growth, and infiltration, owing to its ECM-mimicking nano-micro-macro architecture. Osteogenic studies were conducted by preconditioning the cells with 50 ng/mL BMP-2 for 15 minutes, followed by culturing on mNBC scaffolds for up to three weeks. The results showed an early onset and significantly enhanced bone matrix secretion and maturation in the scaffolds seeded with BMP-2 primed cells compared to the unprimed ones. Moreover, mNBC scaffolds alone were able to facilitate the mineralization of cells to some extent. These findings suggest that, with the aid of 'osteoinduction' from low dose BMP-2 priming of stem cells and 'osteoconduction' from nano-macro/micro topography of mNBC scaffolds, a cost-effective bone tissue engineering strategy can be designed for quick and excellent *in vivo* osseointegration.

## 1. Introduction

Bone fractures and injuries that cause large segmental bone defects do not heal naturally. To restore normal function, the damaged bone is replaced either by a complete artificial substitute (metallic or ceramic) or by harvesting the bone tissue from different areas of the same (autografting) or different person (allografting) or another species (xenografting). However, these practices face significant limitations due to the inadequate supply of donor tissue, pathogen transmission, graft rejection, surgeries at multiple sites, and poor osseointegration [1].

Nowadays, there is a paradigm shift in the practices of bone repair involving tissue engineering, which includes a 3-D biomaterial construct (known as scaffold), stem cells/stem cell-derived progenitor cells, and cellular signals (i.e. growth factors) to make biosynthetic bone grafts for efficient regeneration of damaged or fractured bones [1,2]. The scaffold is one of the key components of tissue engineering that serves as a template to support cell adhesion, proliferation, migration, and differentiation. This scaffold must be biocompatible and should mimic the topography and spatial structures of native extracellular matrices (ECMs) to provide the cues for regulating cell fate towards particular tissue type [3]. Therefore, significant attention is being paid to the materials and fabrication methods for developing ECM-mimicking scaffolds. To date, a number of scaffolds have been developed for bone tissue engineering using various natural and synthetic polymers by distinct techniques such as gas foaming, freeze-drying, particulate leaching, solvent casting, phase separation, electrospinning, etc. [1,4–7]. Amidst various techniques, electrospinning is considered superior as it fabricates nanostructured scaffolds that structurally emulate the natural ECM, but with the inherent disadvantage of limited scaffold thickness and pore size, which restrict three-dimensional cellular growth, resulting into suboptimal outcomes [8,9].

Bacterial cellulose (BC), a polymer produced by a group of acetic acid bacteria owns a natural 3-D nanofibrils network. Meanwhile, it displays admirable mechanical properties (Tensile strength: 200–300 MPa; Young's modulus up to 15 GPa; Compressive strength up to 5 MPa) with excellent biocompatibility, gas and fluid exchange capability, *in situ* moldability, and high water holding capacity [10–13]; making BC a superb scaffolding candidate for tissue engineering. However, the major downside of this naturally woven nanostructured scaffold is inadequate pore size ensuing from tightly packed BC nanofiber layers, which is hard for cells to penetrate [11]; thus hinders tissue in-growth and constrains its practical usage as a tissue engineering scaffold. It has been documented that pore size and pore-interconnectivity are crucial parameters in the development of scaffolds for tissue engineering because they allow cells to migrate and proliferate to the core of the implant as well as permit efficient diffusion of nutrients/metabolic waste, vascularization and remodeling to facilitate host tissue integration upon implantation [3,4]. Thus, introducing macro/micro porosity with apt interconnectivity in nanofibrous BC would lead to an ECM mimicking nano-micro-macro topography, which may suit well for its application in tissue engineering.

Further, growth factors (GFs) play a vital role in tissue regeneration, thus, there are efforts to combine osteoconductive scaffolds with osteoinductive GF(s) for greater outcomes [3,14,15]. However, the covalent binding of GFs to scaffolds is time and cost intensive; in addition, due to arduousness in controlling the modification site, it may also obstruct the active sites of protein and hence impedes GF bioactivity [3,16]. On the other hand, non-covalent incorporation of GF(s) into scaffolds leads to an initial uncontrolled burst release, resulting into supraphysiological levels of GFs in the surrounding environment, which causes unwanted systemic abnormalities and local side effects such as ectopic bone formation, osteoclast-mediated bone resorption, mineral deposition in muscle tissues and inappropriate adipogenesis [3,17].

Preconditioning (using growth factors, pharmacological agents, etc.) approaches in stem cell therapy is an emerging area of research, and several growing reports have publicized enhanced regenerative and repair potentials of preconditioned cells [18–21]. Thus, preconditioning of stem cells with appropriate GF(s) could prove to be a better strategy than GF(s) grafting on the scaffolds for bone tissue engineering applications, as it would minimize the side effects and shortcomings associated with the GFs grafting on the scaffolds as well as the treatment would be cost-effective.

Considering all, the present study explores the potential of 3-D macro/microporous nanofibrous bacterial cellulose (mNBC) scaffolds seeded with BMP-2 preconditioned murine mesenchymal stem cells for bone tissue engineering to repair bone defects and nonunion

## 2. Materials and Methods

### 2.1 Chemicals, culture media, and cells

Dulbecco's Modified Eagle's medium (DMEM), fetal bovine serum (FBS), antibiotic-antimycotic solution, ascorbic acid, β-glycerophosphate, dexamethasone, bone morphogenetic protein-2 (BMP-2), 3-(4,5-dimethylthiazol-2-yl)-2,5-diphenyl tetrazolium bromide (MTT), acridine orange (AO), ethidium bromide (EtBr), alizarin red and lysozyme were purchased from Himedia Laboratories, India. Hexamethyldisilazane (HMDS), phalloidin-FITC (fluorescein isothiocyanate labeled phalloidin), and 4′,6-diamidino-2-phenylindole (DAPI) were acquired from Sigma Aldrich, USA. Murine mesenchymal stem cells (C3H10T1/2 cells) were obtained from National Centre for Cell Science, Pune, India. All the reagents were used as received unless specified otherwise.

### 2.2 Preparation and characterization of 3-D mNBC scaffolds

*Komagataeibacter europaeus* SGP37, previously isolated in our laboratory [22] was used for BC production. BC pellicles were produced by cultivating the bacteria in modified Hestrin-Schramm (HS) medium and purified by boiling in 0.5 N NaOH, as reported earlier by our group [22]. For the fabrication of 3-D mNBC scaffolds, emulsion freeze-drying method described by Xiong et al. [23], was adopted with minor modifications. Briefly, wet BC pellicles (containing refined nanofibrous network) were pulverized to a fine paste using a blender to prepare 2 wt% BC emulsion.  The emulsion was then poured into desired shape molds, frozen at -80 °C for 12 h and lyophilized in a freeze dryer to sublimate frozen water, leaving a highly porous nanofibrous 3-D matrix. The wt% of BC in the emulsion can be optimized for the desired porosity and pore size of the scaffolds.

The microarchitecture of mNBC scaffolds was examined by field emission scanning electron microscopy (FE-SEM) (FEI Quanta 200 FEG, Netherlands) at an accelerated voltage of 15 kV, after coating the scaffolds with gold for 90 s.

X-ray diffraction (XRD) analysis of mNBC scaffolds was carried out by Bruker AXS D8 Advance powder X-ray diffractometer (Karlsruhe, Germany) using Ni-filtered Cu-Kα radiation ($\lambda$ = 1.54 Å). The data were collected in the range of 10-40° 2θ at 2° $min^{-1}$ scan rate. The crystallinity index of the scaffolds and the allomorphic category of cellulose (Iα or Iβ form) after scaffold preparation, were determined as per the formulae described previously [22].

Thermogravimetric analysis (TGA) of the scaffolds was performed using Thermogravimetric/Differential Thermal Analyzer (EXSTAR TG/DTA 6300, Hitachi, Tokyo, Japan) over a temperature range of 25-600 °C at 10 °C $min^{-1}$ scan rate (sample mass ~10.5 mg), under a nitrogen atmosphere with $N_2$ flow rate of 200 mL $min^{-1}$.

### 2.3 Degradation behavior of the scaffolds

Degradation behavior of the mNBC scaffolds was observed in (i) phosphate-buffered saline (PBS) and (ii) phosphate-buffered saline containing 0.2 % lysozyme for 2 months. In brief, the scaffolds were incubated in the above-mentioned solutions at 37 °C. At predetermined time junctures (i.e. 15, 30, and 60 days), the scaffolds were removed, washed thrice with deionized water to remove salts, immersed in 100% ethanol for 2 h, which was then followed by drying at room temperature (RT) to constant weight. The weight loss of the scaffolds (as a measure of degradation) was observed by weighing the dried samples and any microstructural changes in the scaffold's morphology were examined by FE-SEM.

## 2.4 Cell culture

Murine mesenchymal stem cells (C3H10T1/2 cells) were cultured in DMEM high glucose medium supplemented with L-glutamine, 10% FBS, 1% antibiotic-antimycotic solution (containing 10,000 U Penicillin, 10mg Streptomycin and 25µg Amphotericin B per mL) and propagated at 37 °C in a humidified atmosphere of 5% $CO_2$. Once the cells reached 80-85% confluence, they were harvested using trypsin-EDTA solution, resuspended in fresh culture medium at desired density and used for seeding the scaffolds.

## 2.5 Cell seeding and culture on 3-D mNBC scaffolds

Before cell seeding, the scaffolds were sterilized by autoclaving at 121 °C for 20 min, followed by immersion in 70% ethanol under UV for 2 h. Subsequently, the scaffolds were washed twice with sterile PBS and submerged in cell culture medium for 12 h at 37 °C. After incubation, the extraneous culture medium (*i.e.* unsoaked medium) was removed and cells were seeded onto the scaffolds at a seeding density of 30,000 cells/$cm^3$ (except for osteogenic studies where cell seeding density was kept 60,000 cells/$cm^3$ to attain an early confluence), as shown in Table 1.

Following cell seeding, the scaffolds were kept at 37 °C under a humidified atmosphere with 5% $CO_2$ for initial cell attachment and the culture medium was added after 4 h of incubation to prevent unnecessary cell migration to the bottom of the plate. The plates were again placed in the incubator for the duration of the experiments and the culture medium was replenished at each alternate day.

## 2.6 Cell attachment

For cell attachment studies, cell-seeded scaffolds were cultured for 12 h, washed twice with PBS and fixed using 4% formaldehyde for 30 min. After further washing with PBS, the scaffolds were dehydrated using a graded series of ethanol solutions (10%, 20%, 30%, 50%, 70%, 90%, 100%(2x) for 5 min at each step), immersed in 100% HMDS and left to dry at RT. The dried cell-seeded scaffolds were sputter-coated with gold for 90 s and observed under FE-SEM (Carl Zeiss, Ultra plus, Germany).

## 2.7 Cell proliferation and viability

### *2.7.1 MTT assay*

For this, the cell-seeded scaffolds were cultured for up to one week and MTT assay was performed at predetermined time points (1, 4, and 7 days post-seeding) according to the protocol [24,25] with slight modifications. In brief, MTT solution was added to the scaffold-containing wells at a final concentration of 0.5 mg/mL at specified time points, and the plate was incubated in the dark for 4 h at 37 °C under $CO_2$ environment. Post incubation, the solution in the wells was removed and the scaffolds were crushed briefly. Subsequently, DMSO was added to the scaffold (mushed) containing wells and the plate was kept under shaking (100 rpm) at 37 °C for 1 h to dissolve the formazan crystals. Following this, the solution of the wells along with the mushed part of the scaffold was taken out and centrifuged; the supernatant obtained was then read at 570 nm. The entire procedure was followed under darkness. Scaffolds that were not seeded with the cells were taken as a control to ignore the non-specific adsorption of MTT to the scaffolds. Besides quantitative measurements, visual and microscopic images of the formazan crystals formed over the scaffolds were also obtained using a normal digital camera and a CCD camera equipped with an inverted light microscope (Zeiss, Axiovert 25, Germany), respectively.

### *2.7.2 Live/dead cell assay*

C3H10T1/2 cells were grown on mNBC scaffolds for 1, 4 and 7 days. At defined time points, the scaffolds were taken out, rinsed gently with PBS, and stained with equal volumes of acridine orange and ethidium bromide at a final concentration of 5 μg/mL for 30 s. Scaffolds were then visualized and imaged under an inverted fluorescence microscope (Zeiss, Axiovert 25, Germany) to assess cell viability and proliferation.

## 2.8 Cell infiltration

At 1, 4, and 7 days of culture, the cell-seeded scaffolds (48-well sized, area ~9 $mm^2$, thickness ~4 mm) were removed from tissue culture plate, washed twice with PBS, and fixed in 4% formaldehyde for 30 min at RT. Post fixation, paraffin embedding of the scaffolds was done according to the protocol mentioned elsewhere with minor modifications [26]. Briefly, the scaffolds were dehydrated by escalating grades of ethanol (10%, 20%, 30%, 50%, for 10 min at each step and with 70%, 90%, 100%(2x) for 60 min at each step, at RT), treated with xylene (100% for 60 min(2x) at RT) and infiltrated with paraffin wax at 60 °C for 60 min(2x). The scaffolds were then embedded in the wax and cut transversely into ~0.5 mm thin slices using a microtome. Afterward, the slices were dewaxed using xylene, rehydrated using decreasing concentrations of ethanol (100%(2x), 90%, 70%, 50%, $H_2O$(2x) for 5 min at each step at RT), permeabilized with 0.1% Triton-X100 for 20 min and then stained with DAPI (0.3 μg/mL) for 20 min. The sections were then visualized and imaged under an inverted fluorescence microscope.

### 2.9 Cell morphology and cytoskeleton arrangement

Fluorescent staining of actin filaments was done to examine the morphology and cytoskeleton arrangement of cells grown over mNBC scaffolds. Briefly, 3-D cell-scaffold constructs were washed thrice with PBS and fixed using 4% formaldehyde for 30 min at RT. Following further washing with PBS, the cells were permeabilized using 0.1% Triton X-100 for 15 min at RT and stained with phalloidin-FITC (50 μg/mL) for 4h at RT under darkness. The constructs were again washed with PBS (3x) and counterstained with DAPI for 20 min at RT. Subsequently, the constructs were washed several times with PBS to remove the unbound dye and the cells were visualized and imaged using Confocal Laser Scanning Microscope (LSM 780, Carl Zeiss, Germany). Stacks of confocal images were obtained by optical slicing of the scaffolds in Z-direction from top to bottom with 10 μm slice thickness up to 100 μm via Z-stack function of confocal microscopy.

### 2.10 Osteogenic studies

For osteogenic studies, four experimental groups of cell-scaffold constructs (Fig. 1) were maintained:

(i) Group PM: Cells seeded mNBC scaffolds cultured in proliferation medium

(ii) Group PMB: BMP-2 preconditioned-cells seeded mNBC scaffolds cultured in proliferation medium

(iii) Group OM: Cells seeded mNBC scaffolds cultured in osteogenic medium

(iv) Group OMB: BMP-2 preconditioned-cells seeded mNBC scaffolds cultured in osteogenic medium

Prior to seeding, C3H10T1/2 cells were divided into two batches. One batch of cells was preconditioned with BMP-2 by incubating the cells (1,00,000 cells/ml) in proliferation medium (DMEM + 10% FBS) containing 50 ng/mL BMP-2 for 15 min, at 37 °C under $CO_2$ atmosphere. The other batch of cells was used as such without any kind of treatment. Scaffolds (12-well sized, area ~35 $mm^2$, thickness ~9 mm) were then seeded with the respective batch of cells at seeding density of 60,000 cells/$cm^3$ ( i.e. 2,00,000 cells/scaffold) as depicted in Fig.1, and cultured in proliferation medium for two days at 37 °C under a humidified atmosphere of 5% $CO_2$. After two days, the culture medium was replaced with osteogenic media (DMEM, 10% FBS, 10nM dexamethasone, 50 μg/ml ascorbic acid, and 10mM β-glycerophosphate) in groups OM and OMB, while groups PM and PMB were provided with proliferation media up to the duration of the culture. The culture media were changed every second day.

### 2.11 Evaluation of mineralization/calcification

#### *2.11.1 Alizarin red S (ARS) staining*

For ARS staining, the cell-scaffold constructs were washed twice with PBS, fixed in 4% formaldehyde for 30 min, and then rinsed thrice with deionized water followed by staining in ARS solution (1%, pH- 4.3) for 20 min at RT. Excess stain was removed by submerging the scaffolds in deionized water overnight. The scaffolds were then dipped 20 times in acetone, 20 times in acetone-xylene (1:1) solution, followed by clearing in 100% xylene for 15 min. The constructs were photographed using a normal digital camera and the microscopic images were obtained using a light microscope equipped with a CCD camera.

The quantification of mineralization was performed using a colorimetric method [27]. Briefly, after clearing with xylene, the scaffolds were air-dried and added with 10% acetic acid (and crushed to some extent) followed by incubation at RT for 30 min under shaking to extract calcium bound ARS from the scaffolds. The mix obtained was transferred to a tube, vortexed for 1 min, and heated to 85°C for 10 min, then ice-cooled for 5 min. The slurry was then centrifuged and the collected supernatant was mixed with 10% $NH_4OH$ at a ratio of 10:4, which was then followed by absorbance measurement at 405 nm.

#### *2.11.2 Visualization of Ca-P deposits under FE-SEM-EDS*

After 21 days of culture, the cell-scaffold constructs were processed for FE-SEM-EDS analysis according to the protocol detailed in Section 2.6. The samples were then examined under FE-SEM (FEI Quanta 200 FEG, Netherlands), equipped with energy-dispersive x-ray spectroscopy (EDS) (Quanta 200 FEG/EDS, USA, EDAX TEAM Software), to visualize, map, and detect mineral deposits (Ca-P) over mNBC scaffolds grown under various groups.

### 2.12 Statistical analysis

Quantitative data are presented as mean ± standard deviation. Statistical comparisons were made using ANOVA followed by Tukey's post hoc test using Graph Pad Prism 6. The statistical significance was determined at $p \leq 0.05$.

## 3. Results and Discussion

### 3.1 Physicochemical characterization of 3-D mNBC scaffolds

#### *3.1.1 Macro and micro morphologies of 3-D mNBC scaffolds*

Macro/microporous-nanofibrous scaffolds are attracting widespread interest among researchers [28–30] due to their topological similarities with the natural ECM that facilitates high degree of cell adhesion and growth. Fig. 2 (a-c) depicts the macroscopic morphologies of the prepared 3-D mNBC scaffolds. The scaffolds represented are ~ 9 mm in thickness and ~ 18 mm in diameter. However, the size and shape of the scaffolds could be controlled by choosing the molds of specific dimensions to meet the desired size and shape requirements of the tissue. The micromorphology of the scaffolds was examined through FE-SEM (Fig. 2 d,e), which demonstrated that mNBC scaffolds had a highly porous microarchitecture compared to the native BC membrane (Figure S1) and displayed an irregular open-pore geometry with interconnected pore configuration formed by macro-pores (>100 µm, predominantly), micro-pores (<100 µm) and nano-pores (<100 nm).

The pore size and pore architecture are the crucial aspects to be considered when designing scaffolds for tissue engineering, as they guide biological processes towards regeneration. However, there are conflicting reports on the optimal pore size of scaffolds for bone tissue engineering. Scaffolds with mean pore sizes ranging from 20-1500 µm have been applied for this purpose and it has been advocated that pores >300 µm enhance the osteogenic potential of scaffolds [28,31]. However, some reports have shown that microporosity (pore size <100 µm) plays a significant role in enhancing the osteogenic potential of scaffolds [32,33]. In this regard, it has been put forward that micro-pores (<100 µm) play a beneficial role in initial cell adhesion due to their greater surface area; but, due to their smaller aperture, they enable cell aggregation and, limit cell infiltration and migration [34]. This restricts vascularization and contributes to a hypoxic environment (pro-chondrogenic) within the scaffold that favors endochondral ossification (chondrogenesis before osteogenesis) [35–38]. Macro-pores, on the other hand, are less conducive to cell adhesion and habitation, but encourages cell infiltration and migration which facilitates vascularization and high oxygenation; thus induces intramembranous ossification (i.e. bone formation without preceding cartilage formation), as well as enables better host tissue integration [34,35,37,38]. Besides pore size, pore interconnectivity is another essential factor to be considered, as pore interconnections facilitate cell distribution, integration with the host tissue, and capillary ingrowth [39].

Thus, the nanofibrous topography, pore size heterogeneity and pores interconnectivity of the prepared mNBC scaffolds may be conducive to both, (i) initial cell adhesion (due to the larger surface area provided by micro-pores and nanofibers) and later, (ii) cell infiltration and migration (owing to the macro-pores); which may lead efficient bone regeneration, vascularization, and host tissue integration upon implantation. However, the scaffold's pore size, pores homo/heterogeneity, pore morphology, pore configuration, fiber positioning and orientation also play an important role in determining the ultimate mechanical properties of the scaffold, and that need to be investigated further and should be accounted for better tuning of structural and mechanical cues, for successful bone repair.

### *3.1.2 X-Ray diffraction and thermogravimetric analyses of 3-D mNBC scaffolds*

X-ray diffraction analysis of mNBC scaffolds was carried out to observe any physical changes in the polymer after scaffold preparation. As shown in Fig. 2 f, the diffractogram showed three characteristic peaks at 2θ = 14.6, 16.2 and 22.2°, corresponding to $(1\bar{1}0)$, (110) and (200) crystallographic planes of the cellulose lattice, respectively [22,40], which was similar to the XRD profile of the native BC membrane (Figure S2). However, the crystallinity index (CrI) of the cellulose got decreased from 87.7 % (native BC membrane) to 54.66 % after scaffold preparation and the allomorphic category of the cellulose also changed from Cellulose Iα (triclinic lattice structure) to Cellulose Iβ (monoclinic lattice structure). This could be due to the disruption of cellulose chain assembly during the crushing process used for the preparation of scaffolds.

Thermogravimetric analysis of mNBC scaffolds (Fig. 2e) also showed a similar TG-DTG profile to that obtained with native BC membrane (Figure S3) with three characteristic phases of weight loss [22]. However, the degradation of mNBC scaffolds started at ~285 °C and 33.19 % mass was remained at 350 °C, while the native BC membrane was stable up to ~295 °C and 43.69 % residual mass was present at 350 °C. Decreased thermal stability of mNBC scaffolds could be attributed to their lowered crystallinity index, as the thermal degradation behavior is reported to be persuaded by the crystallinity and orientation of the BC nanofibers [40].

**3.2 Degradation behavior of the scaffolds**

An ideal scaffold is required either to degrade or to be reabsorbed by the body after tissue regeneration. Being a polysaccharide, BC is unlikely to get affected by the proteases. Hence, the *in vitro* degradation behavior of mNBC scaffolds was examined in PBS and PBS containing lysozyme solutions, as lysozyme is present in almost all body fluids [41]. Although lysozyme mainly breaks the β-1,4 glycosidic linkage between the NAM and NAG units of peptidoglycan, it may also affect the β-1,4 glycosidic linkage of cellulose to some extent. Thus, mNBC scaffolds were incubated in PBS (pH 7.4) and PBS containing lysozyme (0.2%, pH 7.4) solutions separately, at 37 °C for 15, 30, and 60 days to assess the weight loss in scaffolds due to dissolution and degradation. No significant change in the weight of the scaffolds was noted in either of the solutions even after 60 days, as well as no deterioration in the scaffold's microarchitecture was noticed either, except little crumpling (Figure S4), which indicates that the time needed for the degradation of mNBC scaffolds may be longer than the observation period.

A study by Martson et al.  in a rat model reported that the cellulose sponges did not degrade completely even after 60 weeks. Although they were totally filled up with the connective tissues after 8 weeks of implantation, but the emergence of cracks and fissures, and slackening of the pore walls of the cellulose sponges were observed only after 16 weeks, hence they regarded it as a slowly degradable implantation material [42]. Nevertheless, due to the slow degradation behavior of mNBC scaffolds, they could be explored for the applications, which takes longer time to heal such as larger bone defects, bone tumors, and bone non-unions that take months to years to heal, as shown in a recent analysis of clinical trials using various kind of scaffolds [43].

Further, since cellulose is generally degraded in nature by microbial enzymes through hydrolase attack on the β-1,4 linkages, but these enzymes are not present in the mammals. In this case, cellulose degradation is likely to occur by an amalgamation of chemical, biological and mechanical processes. In other words, its degradation is controlled by several factors, such as its crystallinity, aggregation state, surface area, shape and morphology of the scaffolds, pH, hydrophilicity, and the accessibility of its hydrolytic unstable bonds to the body fluids [42,44]. In fact, a phenomenal decrease in crystallinity index was seen to be associated with rapid degradation process of BC due to the transformation of crystalline region into amorphous region [45]. Thus, the highly porous interconnected geometry of the scaffolds and the reduced crystallinity of the cellulose after scaffold preparation (as mentioned in Section 3.1.2) may be advantageous for its post-implantation degradation. Apart from this, additional improvements in the BC can be made to tailor the degradation rate of mNBC scaffolds. For instance, periodate oxidation of BC has the potential to increase its degradability by opening of the glucopyranose rings and disrupting its higher ordered structure [46]. Incorporating cellulase enzyme within the scaffolds is another strategy to regulate the degradation rate of BC [45], which can be considered to tune the degradation rate of mNBC scaffolds with the growth rate of neo bone tissue.

## 3.3 Cell studies

### *3.3.1 Cell attachment, proliferation, viability and infiltration*

Cell behavior, such as adhesion, proliferation, spreading, and infiltration represents the initial phase of cell-scaffold communication which subsequently impacts the further events viz. differentiation and mineralization [47]. Although the biocompatibility of bacterial cellulose is well known, but the source of BC synthesis, the method used for BC scaffolds preparation, the scaffold characteristics and the cell type being used, make the preliminary cell–scaffold interaction studies essential for any further studies to begin with.

In this regard, Fig. 3 displays the representative electron microscopic images of murine MSCs (C3H10T1/2 cells) adhered to mNBC scaffolds 12 h post-seeding. The images showed that the cells adhered well to the scaffold and maintained an extended fibroblast-like morphology. High-magnification images revealed that the cells adhered to the scaffold with their pseudopodium anchored on the walls of BC nanofibers, an indication of typical cell attachment and growth process.

After adhesion, cells enter into the proliferation phase, hence, C3H10T1/2 cell proliferation on the mNBC scaffolds was investigated by MTT assay and the findings are summarized in Fig. 4a. MTT staining of cell-seeded scaffolds (Fig. 4a(i)) at various time points (days 1, 4, and 7) revealed that the cells were metabolically active (as indicated by purple color intensity) and gradually migrated throughout the scaffold during the time. Quantified data of MTT staining (Fig. 4a(ii)) showed a significant increase ($P < 0.001$) in the metabolic activity as a function of time for the duration of the experiment and represented a high cell viability and proliferation, indicating the biocompatible nature of mNBC scaffolds. To verify the results obtained from MTT assay, the viability and proliferation of C3H10T1/2 cells on mNBC scaffolds was further assessed by live/dead cell staining. The fluorescent images showing live cells in green and dead cells in red (Fig. 4b), reaffirmed the biocompatibility of mNBC scaffolds with hardly any detectable cell death. The cells continued to proliferate throughout the scaffold, as the results showed a rise in cell number over time, however, due to the 3-D nature of the scaffolds, the scattered cells could not be clearly imaged at a single focus but were visible on different planes. These findings are attributed to the highly porous interconnected geometry of mNBC scaffolds due to which the diffusion of nutrients and waste across the scaffold was ensured, rendering the cells to proliferate and be metabolically active.

Rapidly attracting, recruiting, and dispersing the surrounding cells through the 3D matrix is one of the key requisites for the success of implantable tissue engineering scaffolds [48]. Cell ingress into mNBC scaffolds has thus been investigated during culture for up to 7 days by DAPI staining of cell nuclei in scaffold cross-sections. As delineated in Fig. 5a, the majority of seeded cells were present at/near the scaffold surface and a very small percentage of cells could be seen into the depth of the scaffold on day 1. However, by day 7, cells had infiltrated, proliferated, and homogeneously disseminated throughout the entire depth of scaffold; indicating the potential of mNBC scaffolds for tissue in-growth as well.

The microscopic morphology of C3H10T1/2 cells on mNBC scaffolds was confirmed using CLSM, where the cells were stained with phalloidin tagged FITC for visualization of cytoskeletal processes and with DAPI to visualize nuclei. The results demonstrated a well spread, elongated, fibroblast-like morphology of C3H10T1/2 cells (Fig. 5b(ii)), consistent with the FE-SEM observation (Fig. 3). The cell infiltration inside the mNBC scaffolds was further confirmed by optical slicing of the scaffolds in Z-direction from top to bottom with 10 μm slice thickness through the Z-stack function of confocal microscopy (Fig. 5b(i)). Reconstructed 3-D projection images of the scaffold on day 4 post-seeding exhibited adequate growth, proliferation, and infiltration of cells, further supporting the fact that the mNBC scaffolds are ideal for tissue engineering applications.

### *3.3.2 Osteogenic studies with BMP-2 preconditioned murine mesenchymal stem cells*

Bone regeneration is regulated by a cascade of molecular factors, however, BMPs (bone morphogenetic proteins) play a critical role in initiating the fracture repair cascade and primarily act by triggering osteogenic differentiation of osteoprogenitors and recruiting MSCs to the injured area [3]. In particular, BMP-2 and BMP-7 are considered the most potent osteoinductive cytokines, which are approved by FDA for the clinical practices [17]. However, the controlled delivery of BMPs to the site of injured bone tissue continues to be an arduous task due to the variable release profiles of BMPs from the carriers [3,17]. For instance, adsorbing BMPs to the implant surface leads to an early, uncontrolled, burst release of GF when exposed to the physiological environment. Immobilizing BMPs to the surface of the implant maintains a sustained presence of GFs; however, due to arduousness in controlling the modification site, covalent binding may block active sites of the protein and thus impedes the bioactivity of GF. Encapsulation and entrapment of BMPs evade the hitches associated with adsorption and immobilization, thus are the most popular way to deliver GFs; but many of these methods expose BMPs to harsh solvents, which may distort the conformational structure of the protein, thereby interferes with GF activity. All these hassles lead to the need for supraphysiological loading of BMPs (0.02-0.4 mg/mL in rodents, 0.75-2.0 mg/mL in nonhuman primates and 1.5-2.0 mg/mL in human clinical practices) in the carrier devices to achieve an efficacious outcome [17,49,50]. These doses, approximately 10 million times higher to that found in the body (pg/ml) have made BMP-based treatments very expensive (~$5,000 or more), and is believed to be responsible for many of the complications seen in BMP-based therapies, such as ectopic bone formation, native bone resorption, mineral deposition in muscle tissue, soft tissue swelling, etc. [17,50,51]. Therefore, novel systems for BMP delivery and the alternative approaches to harness optimal BMP effect at low BMP concentration, are continue to receive attention.

Preconditioning strategies in stem cell therapy are currently catching the attention of researchers as a variety of preconditioning triggers such as sublethal insults (e.g. ischemia, anoxia, hypoxia), growth factors (e.g. SDF-1, ILGF-1, BMP-2), pharmacological agents (e.g. apelin, diazoxide, isoflurane) are found to increase the regenerative and repair potential of stem cells and stem cell-derived progenitor cells [18–21]. Thus, we preconditioned MSCs with a very low dose of BMP-2 (50 ng/mL) for 15 minutes prior to seeding on the scaffolds. Cell-seeded scaffolds were then cultured in complete media with/without osteogenic inducers (β-glycerol-2-phosphate, dexamethasone and ascorbic acid) to determine whether BMP-2 preconditioning could modulate the stem cells behavior towards osteogenic differentiation. Two propositions led us to speculate that preconditioning of cells with BMP-2 could be sufficient to elicit osteogenic responses; (1) BMP-2 activity is mainly required at the initial stages of fracture healing, and (2) due to the short (in minutes) systemic half-life of BMP-2, its activity vanish gradually, even when administered locally in the scaffolds [52–54].

One of the hallmarks of osteogenic differentiation is the formation of extracellular mineralized deposits of calcium and phosphorus salts, in which the anionic matrix molecules bind with $Ca^{2+}$and $PO_4^{3-}$ ions, and thereafter serve as sites for nucleation and growth, leading to calcification [47]. Alizarin Red S staining was used to probe these mineral deposits on various groups of cell-seeded mNBC scaffolds at different time points (7, 14, and 21 days). Fig. 6(a-c) displays the optical and microscopic images of the ARS staining for the cell-seeded mNBC scaffolds and Fig. 6d depicts the respective quantitative data obtained after extracting ARS with 10% acetic acid. Control scaffolds (without cells) incubated in differentiation medium, showed no positive stain (Figure S5), thus ruling out the point of dye absorption/adsorption by the scaffolds.

The results demonstrated a variation in the extent of mineralization with culture time, culture media, and cell preconditioning. At day 7, the scaffolds cultured in osteogenic medium (i.e. Group OM and OMB) exhibited areas of diffuse and nodular mineralization, where group OMB (i.e. seeded with BMP-2 preconditioned MSCs) was slightly more intense than group OM (seeded with unprimed MSCs), however, it did not reach the statistical significance ($P > 0.05$). On the other hand, the scaffolds grown in proliferation medium (i.e. Group PM and PMB) showed a very low intensity diffused staining pattern, with no accountable difference between the groups, with reference to cell preconditioning. The similar trend was observed at day 14, but with relatively higher stain intensity in all the groups. However, some nodular mineralization was noticed in group PMB and the difference in the stain intensity of the scaffolds of group OMB and OM reached statistical significance ($P < 0.01$), suggestive of BMP-2 induced osteogenesis of preconditioned cells. At day 21, the stain intensity was much higher in all the groups, in fact, scaffolds of group OM and OMB completely turned red. Additionally, the stained deposits were witnessed throughout the scaffold depth from top to bottom, implying time dependence of cell mineralization to produce more $Ca^{2+}$ binding sites for ARS. Moreover, the scaffolds seeded with BMP-2 preconditioned cells showed a higher score of stain intensity compared to the scaffolds seeded with unprimed cells, under both proliferation ($P < 0.01$) and osteogenic medium ($P < 0.001$) which further propounds the role of BMP-2 preconditioning in influencing osteogenesis of primed cells. The highest score of stain intensity in group OMB throughout the culture period is obviously attributed to the presence of DAG (dexamethasone/ascorbic acid/β-glycerolphosphate) in the culture medium, as DAGs are reported to enhance BMP-2 induced osteogenesis [55,56]. Interestingly, it was notable that mNBC scaffolds were able to facilitate mineralization of murine MSCs after 3 weeks of culture, even in the absence of osteogenic stimulants (DAG) and BMP-2 preconditioning. This may be due to the topographical characteristics of mNBC scaffolds, which might have provided osteogenic cues to the cells, leading to calcification. Thus, with the aid of BMP-2 primed cells and the intrinsic ability of the mNBC scaffolds to induce osteogenesis, a cost-effective bone tissue engineering strategy can be designed for quick and excellent *in vivo* osseointegration.

A cross confirmation of the mineralized matrix over cell-scaffold constructs was done using FE-SEM followed by EDS, after 21 days of culture. The electron micrographs revealed that scaffolds were covered with the network of cells, extracellular matrix, and globular accretions of ~ 0.5-10 µm, suggestive of calcification (Fig. 7a) [57]. The size of the accretions was larger in the scaffolds of group OMB compared to the other groups, inferring enhanced calcification in group OMB. For verification of these accretive bodies to be of calcium phosphate, EDS mapping was performed (Fig.7b), which confirmed the accumulation of calcium phosphate at respective positions. The quantitative measures of EDS analyses also revealed the presence of calcium and phosphate in all the four groups of cell-scaffold constructs and the intensity of the peaks referring to these elements were found to be in the order

OMB>OM>PMB>PM (Fig.7c). These results also depicted a similar trend in mineralization as obtained with ARS staining, suggesting that BMP-2 preconditioning of stem cells prior to seeding on scaffolds could be of potential benefit in tissue engineering of bone defects.

Next, the cell-scaffold constructs were subjected to phalloidin- FITC and DAPI staining to find out how cells were behaving morphologically after 21 days of culturing on mNBC scaffolds (Fig. 8) as cell shape, cell area and cytoskeleton arrangement provide a unique way to characterize the differentiation directions [58]. The results revealed excellent adhesion and growth of cells on mNBC scaffolds even after 21 days of culture, however, cells looked overlapped and stacked on one another, and were found lining the walls of the scaffold pores. The cytoskeleton staining of cells grown under group PM displayed that some cells attained a broad shaped polygonal morphology while some were still elongated with spindle-shape (Fig. 8a, also refer Figure S6 for individual images of Z- stack). On the other hand, as expected, almost all the cells of group OM attained a broad polygonal morphology with a large increase in cell area (Fig. 8a and Figure S7), suggestive of their commitment towards osteogenesis [59].

Cumulatively, these findings indicate that the combination of BMP-2 primed stem cells and 3-D macro/microporous nanofibrous bacterial cellulose scaffold could be a promising tissue-engineering tactic to repair bone defects and nonunion, for which patient's own MSCs could be isolated, preconditioned with BMP-2, seeded on the scaffold and thereafter implanted at the repair site for quick and efficient bone regeneration. Since a low dose of BMP-2 was used in this approach and that too affects only the cells stimulated *ex vivo*; thus, it would avoid the off-target adverse effects of high dose BMP-2 therapy as well as would cut down the cost of the treatment.

Growing evidence supports increased regenerative potential of preconditioned cells in tissue repair and functional recovery. A study by Knippenberg et al. demonstrated that preconditioning of goat ADMSCs with 10 ng/mL BMP-2 for 15 min triggered the osteogenic differentiation of cells and preconditioning with BMP-7 under same conditions stimulated a chondrogenic phenotype [60]. Lysdahl et al. [19] found that preconditioning hBMMSCs with 20 ng/mL BMP-2 for 15 min increased the proliferation and osteogenic differentiation of cells, while continuous exposure of cells with the same concentration of BMP-2 only increased the proliferation and did not make any difference in osteogenic differentiation. Martinez et al. demonstrated that *ex vivo* preconditioning of mouse BMMSCs with 2nM BMP-2 and 50 ng/mL Wnt3a for 24 hours, followed by seeding on gelatin/CaSO4 scaffolds, cooperatively enhanced the expression of osteogenic markers *in vitro* and bone regeneration in a critical-size calvarial defect mouse model [20]. Muzio et al. [61] demonstrated the role of shock wave preconditioning in modulating the osteogenic properties of osteoblast-like cells seeded on a bioactive scaffold. A study by Watanabe and coworkers showed that preconditioning of rat BMMSCs with 5mM NAC (N-acetyl-L-cysteine) followed by seeding on collagen sponge significantly enhanced bone regeneration in critical-sized rat femur defect by strengthening BMMSCs resistance to apoptosis and senescence induced by oxidative stress during the acute inflammatory phase of injury [62].

Though preconditioning of stem cells or stem cell-derived progenitor cells have shown promising results in tissue repair and functional recovery, however, until now, *in vitro* preconditioning of cells for *in vivo* tissue regeneration has been challenging. Cells preconditioned for a longer period generally fail to integrate with the host, possibly due to the maturity of tissue; on the other hand, shorter preconditioning periods appear not to be sufficient in terms of clinically relevant outcomes [63]. Apart from this, many more hitches are associated with cell preconditioning for *in vivo* tissue regeneration; (1) will preconditioning factors exert adverse effects on cells, (2) will stem cells acquire tumorigenicity after preconditioning, (3) what should be the safe dose of these factors? All these apprehensions need to be clarified before proceeding further in this direction.

## 4. Conclusions

Collectively, 3-D macro/microporous-nanofibrous bacterial cellulose (mNBC) scaffolds were prepared and applied to direct osteogenic differentiation of low dose BMP-2 primed murine mesenchymal stem cells, in order to develop an efficient and cost-effective bone tissue engineering strategy. The ECM-mimicking nano-micro-macro architecture of mNBC scaffolds provided an excellent environment for cell adhesion, growth, infiltration, and to some extent osteodifferentiation also, making it ideal for bone tissue engineering. Osteogenic studies demonstrated significantly enhanced bone matrix secretion and maturation when the scaffolds were seeded with BMP-2 primed cells compared to the unprimed ones and a synergistic effect towards calcification was seen when BMP-2 primed-cells-scaffold constructs were provided with osteogenic stimulants (DAG) during the culture. However, additional studies at molecular level are required to corroborate these findings, as well as further studies to elucidate the underlying mechanisms, including intracellular signal transduction pathways are warranted.

Nevertheless, these exploratory findings suggest that adopting low dose BMP-2 preconditioning of stem cells in a bone tissue engineering strategy may provide a promising solution to alleviate the economic and negative impact of high dose BMP-2 grafting on scaffolds, as well as may offer a paradigm to design strategies for stem cell fate direction in other arenas of regenerative medicine.

## Acknowledgments

SD is grateful to the Department of Biotechnology (DBT), Government of India for providing financial assistance (DBT- JRF/10-11/318) to undertake this work. Thanks to Dr. Ritu Varshney for helping in CLSM image acquisition. Sincere gratitude to Institute Instrumentation Centre, IIT Roorkee, for the various analytical facilities provided.

## Appendix A. Supplementary data

## Tables

**Table 1** Details of the cell seeding on mNBC scaffolds

| Seeding density (cells/cm$^3$) | Scaffold's shape | Scaffold's dimensions (mm) | Average calculated number of cells per scaffold | Seeding cell suspension volume (µl) |
|---|---|---|---|---|
| 30,000 cells/cm$^3$ | Cylindrical (12-well sized) | Area ~35 mm$^2$ Thickness ~9 mm | 1,00,000 cells/scaffold | 200 µl |
| | Cylindrical (24-well sized) | Area ~18 mm$^2$ Thickness ~9 mm | 50,000 cells/scaffold | 100 µl |
| 60,000 cells/cm$^3$ (for differentiation studies) | Cylindrical (12-well sized) | Area ~35 mm$^2$ Thickness ~9 mm | 2,00,000 cells/scaffold | 200 µl |

# Figures

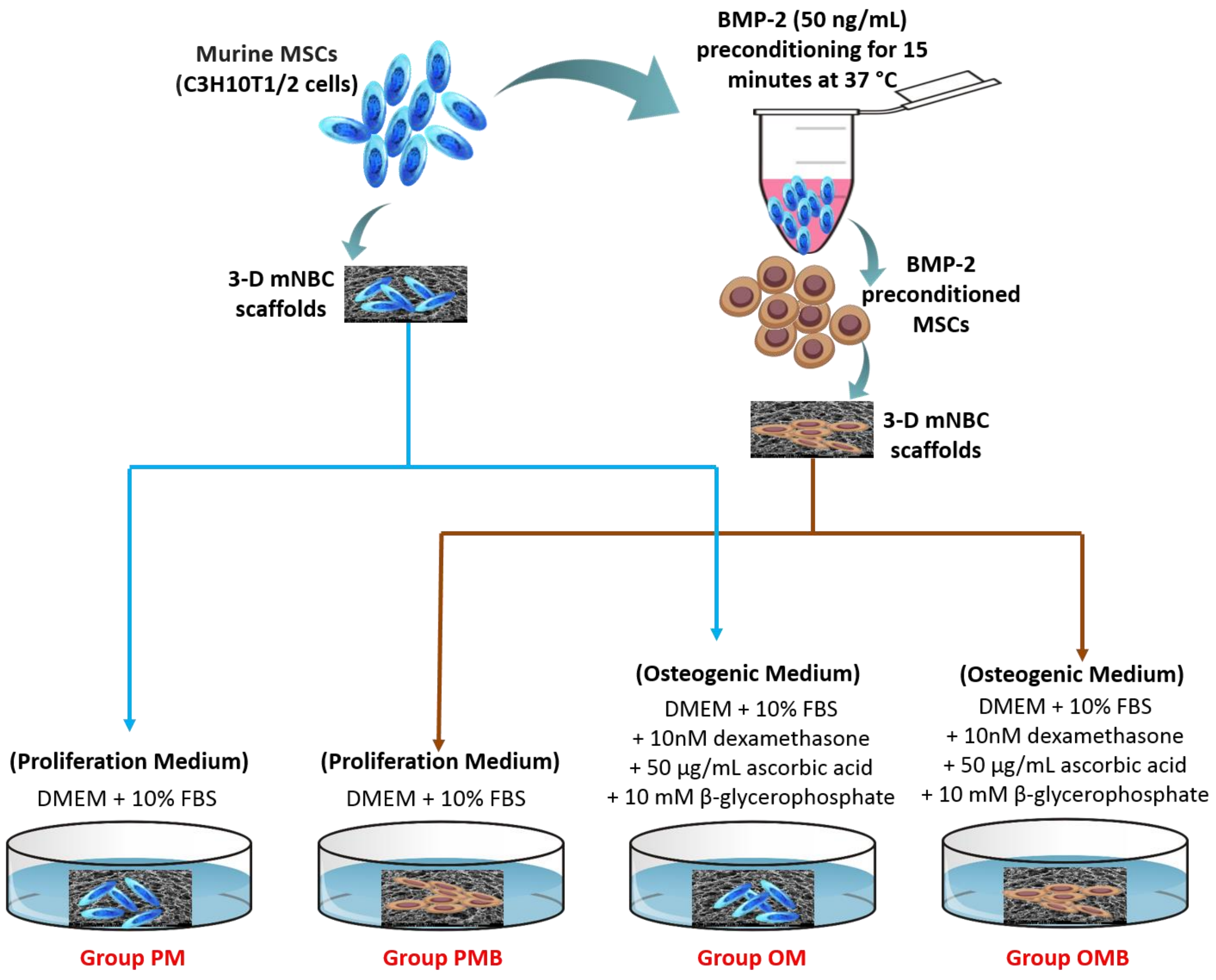

**Fig. 1** Cell culture strategy for osteogenic studies

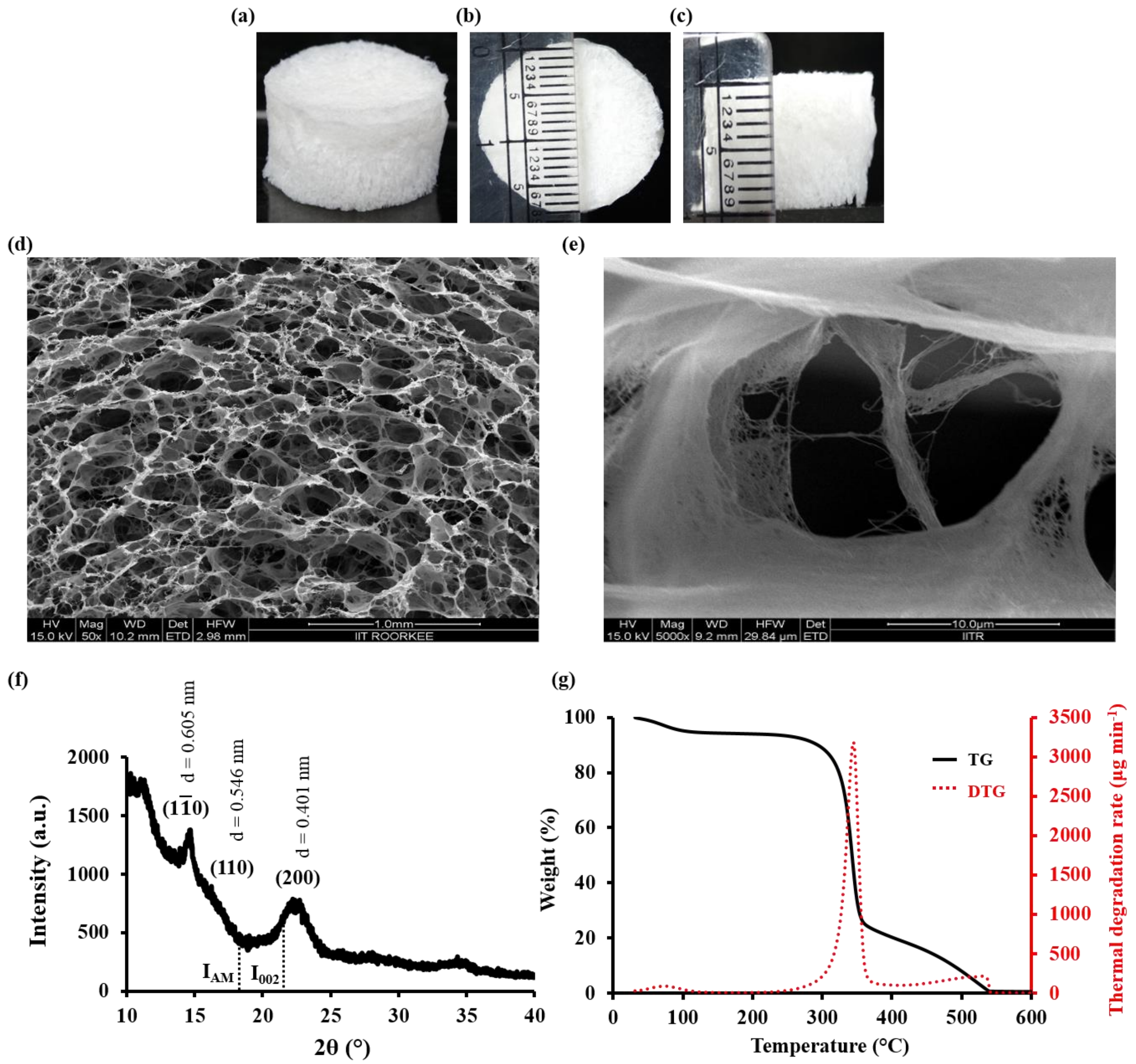

**Fig. 2** (a-c) Visual aspects of mNBC scaffold in terms of its three dimensionality and size; (d-e) microscopic morphology of mNBC scaffold at (d) 50 X and (e) 5k X magnifications showing a macro/microporous-nanofibrous architecture; (f) X-ray diffraction profile and (g) TG-DTG curve of mNBC scaffolds

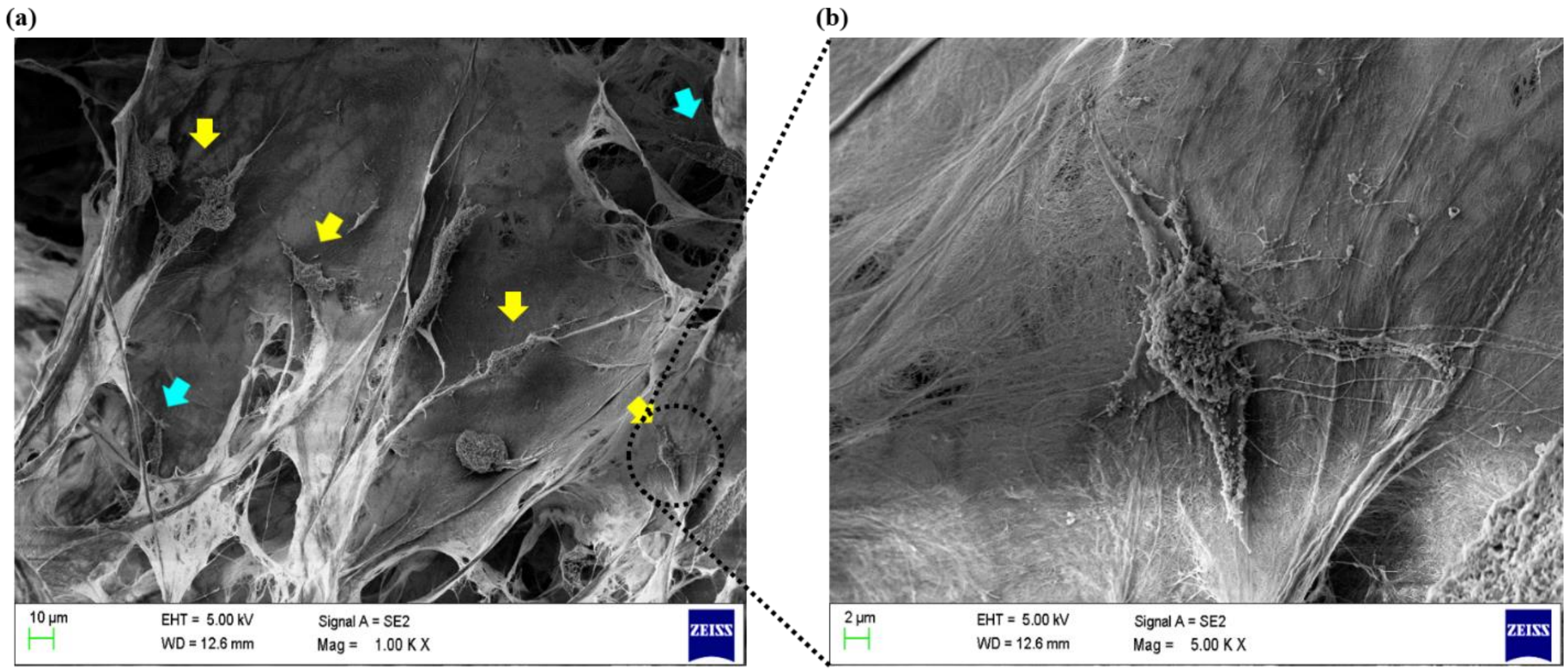

**Fig. 3** FE-SEM micrographs showing C3H10T1/2 cells (murine MSCs) attachment on 3-D mNBC scaffolds at (a) 1kX and (b) 5kX magnifications (cells were located on the surface (yellow arrow) as well as within the pores (cyan arrow) of the scaffold)

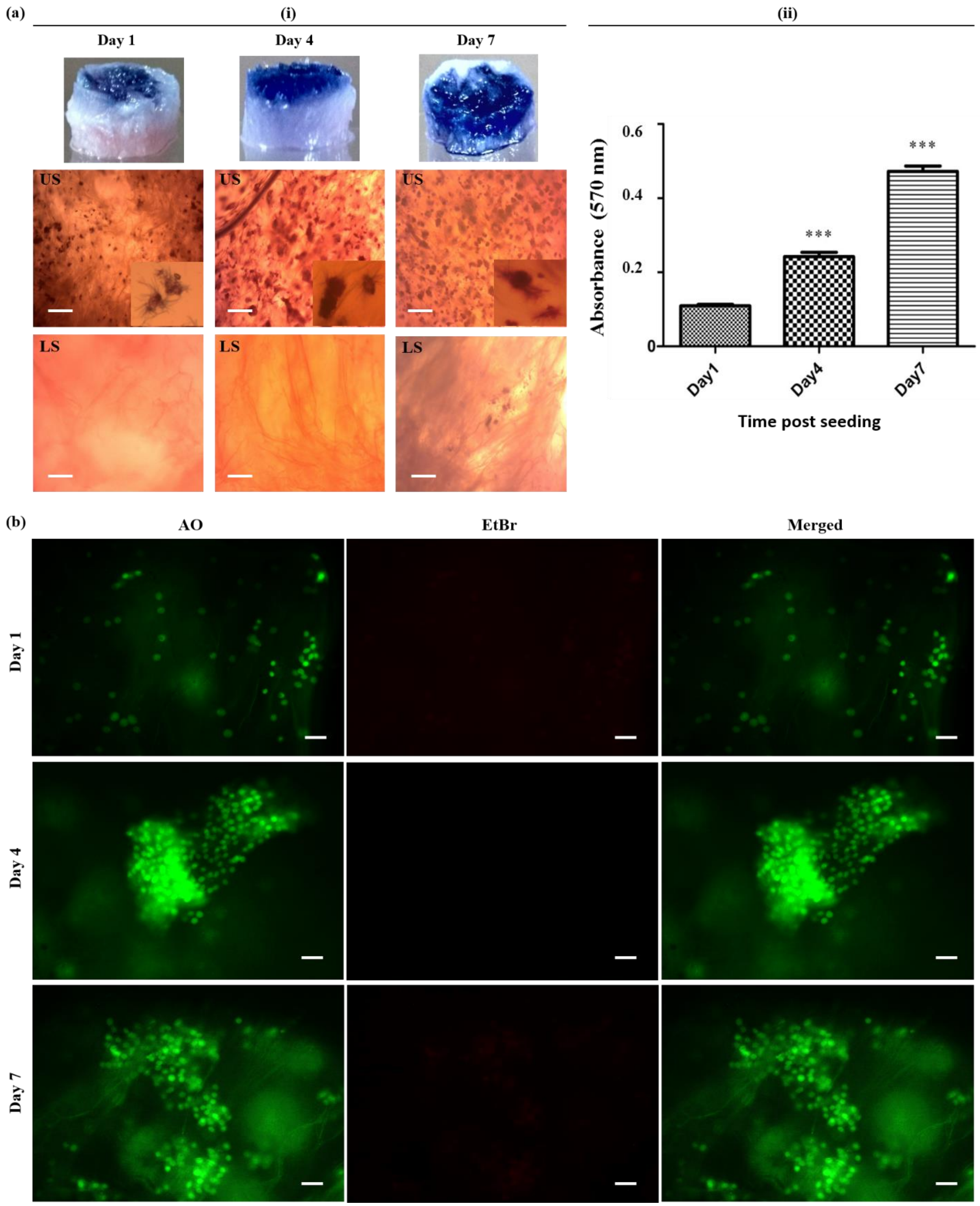

**Fig. 4** (a) MTT assay of C3H10T1/2 cells seeded mNBC scaffolds at days 1, 4, and 7. Panel (i) illustrates the visual and microscopic aspects of the MTT assay (inset displays the magnified view of formazan crystals) showing a rise in purple colour intensity over time due to an increase in metabolically active cells. Panel (ii) depicts the histogram representing quantitative data (n=6; $P < 0.001$) of MTT assay [Scale bar = 100 µm; US: Upper Side; LS: Lower Side of the scaffold]. (b) Live/dead cell staining of C3H10T1/2 cells seeded mNBC scaffolds at days 1, 4, and 7. Live cells were stained green by acridine orange (AO) and dead cells were stained red by ethidium bromide (EtBr) [Scale bar = 50 µm].

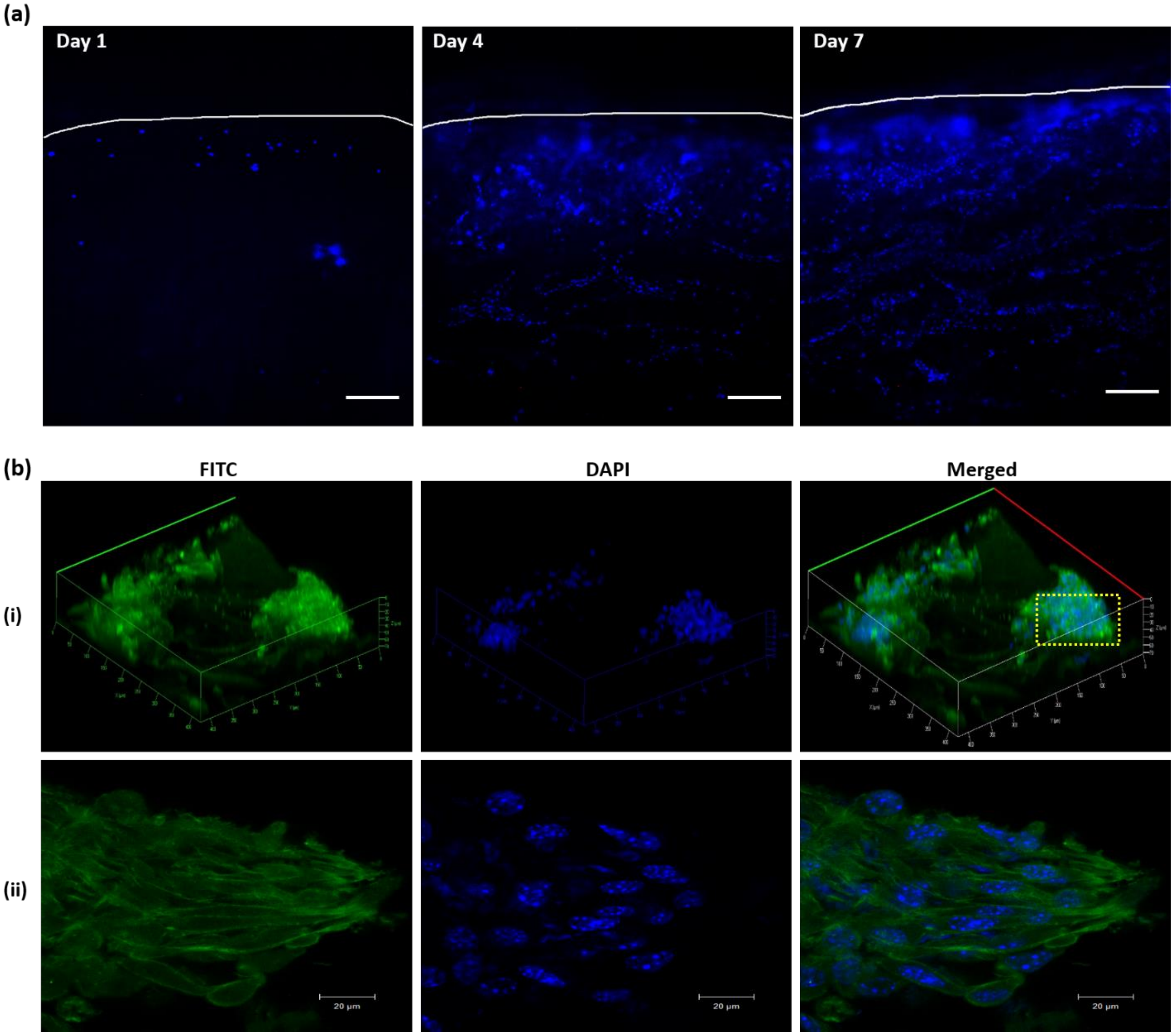

**Fig. 5** (a) C3H10T1/2 cells infiltration into 3-D mNBC scaffolds at days 1, 4, and 7, demonstrated by DAPI staining of cell nuclei in scaffold transverse-sections. The sketched white lines on the fluorescent images are intended to set out the outer edges of the scaffold [Scale bar = 200 µm]. (b) Confocal micrographs of phalloidin-FITC and DAPI stained C3H10T1/2 cells cultured on 3-D mNBC scaffolds: Panel (i), 3-D projection of Z-stack images from surface to bottom of the scaffold (up to 70 µm); Panel (ii), cytoskeletal morphology of C3H10T1/2 cells on mNBC scaffolds (from selected rectangle area).

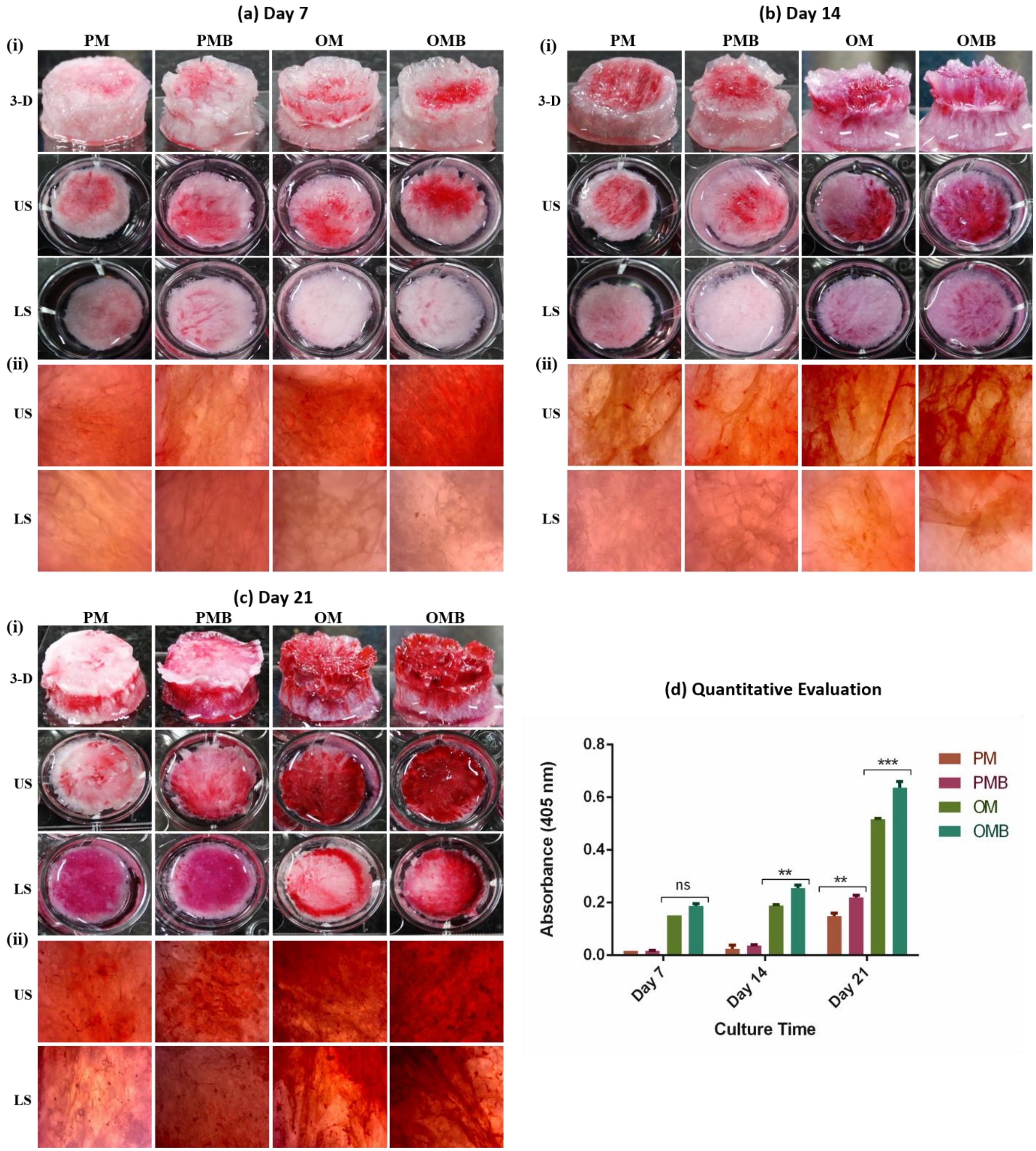

**Fig. 6** Alizarin red staining of C3H10T1/2 cells seeded mNBC scaffolds at (a) day 7, (b) day 14 and (c) day 21. Panel (i) and (ii) shows the visual and microscopic images of ECM mineralization on the scaffolds, respectively. Respective quantitative data (d) obtained after extracting ARS from the stained scaffolds is presented as mean ± SD based on at least triplicate observations from two independent experiments. (PM: proliferation media; PMB: proliferation

media+BMP-2 preconditioning of cells; OM: osteogenic media; OMB: osteogenic media+BMP-2 preconditioning of cells; 3-D: 3-dimensional; US: upper side of the scaffold; LS: lower side of the scaffold)

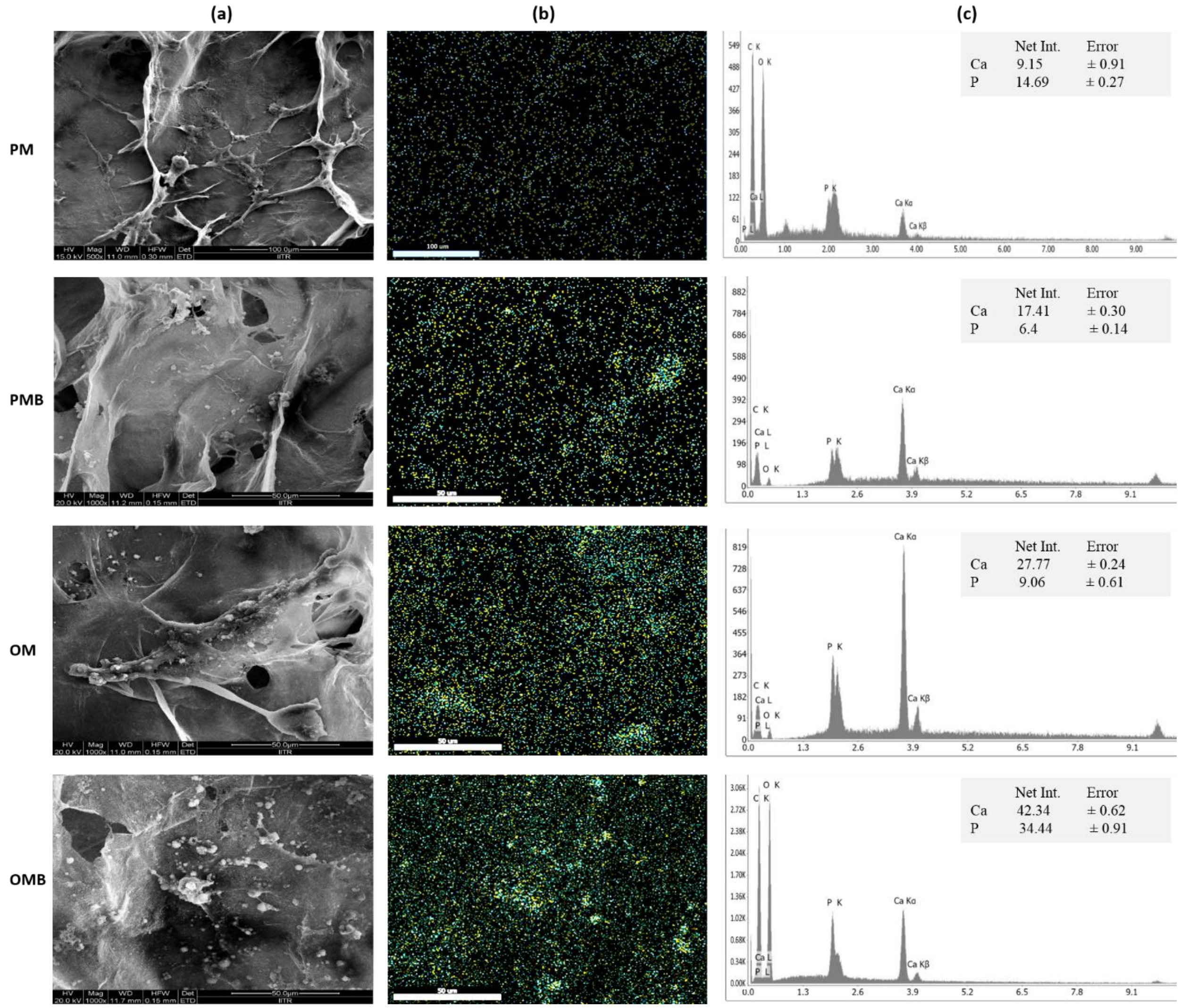

**Fig. 7** FE-SEM and EDS analyses of C3H10T1/2 cells-seeded mNBC scaffolds after 21 days of culture: (a) FE-SEM micrographs showing mineralized matrix, (b) EDS mapping confirming the presence of Ca and P in the accretions (Cyan blue dots- Ca; Yellow dots- P), (c) EDS spectra showing quantitative measures of Ca and P. (PM: proliferation media; PMB: proliferation media+BMP-2 preconditioning of cells; OM: osteogenic media; OMB: osteogenic media+BMP-2 preconditioning of cells)

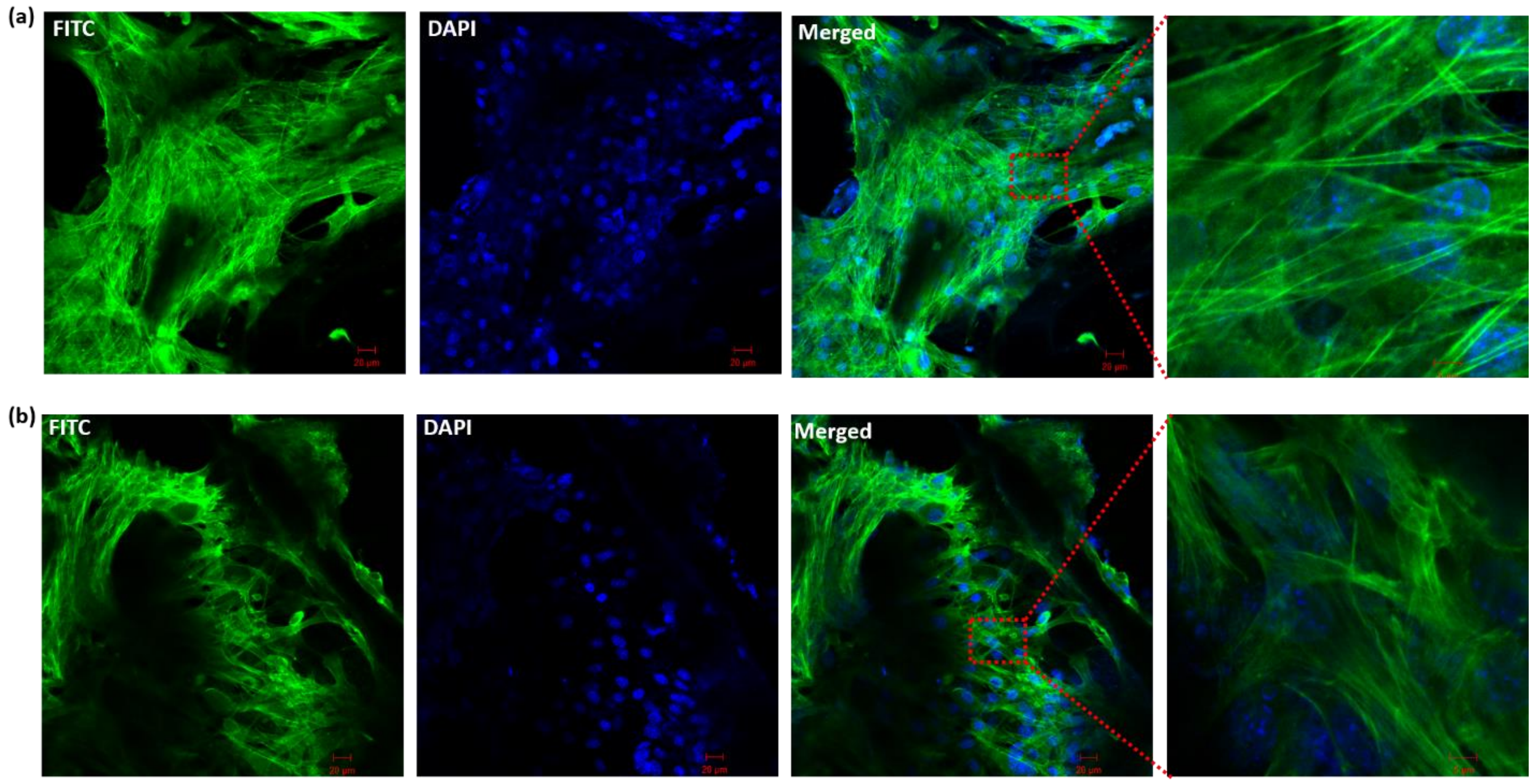

**Fig. 8** Cellular morphology and cytoskeleton arrangement of C3H10T1/2 cells grown on 3-D mNBC scaffolds after 21 days of culture in (a) proliferation (PM) and (b) osteogenic (OM) medium. Actin cytoskeleton is shown in green by fluorescent staining with phalloidin-FITC and nuclei are shown in blue by DAPI staining. (These images are 2-D reconstruction of individual Z-stack images (provided as supplementary data) obtained by the optical slicing of samples in the Z-direction through Z-stack confocal microscopy)

**Annexure A. Supplementary data**

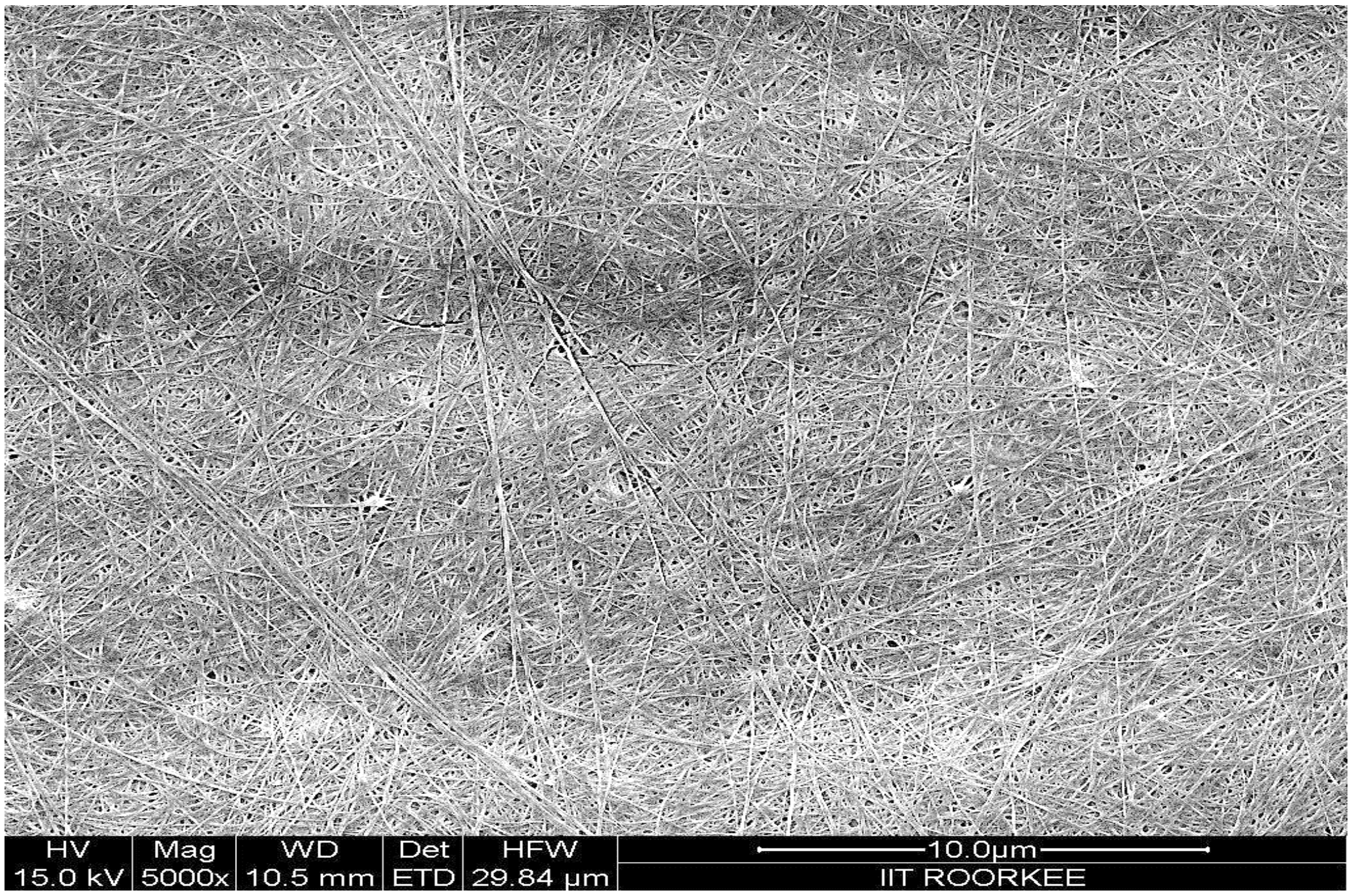

**Fig. S1** Microarchitecture of native BC membrane

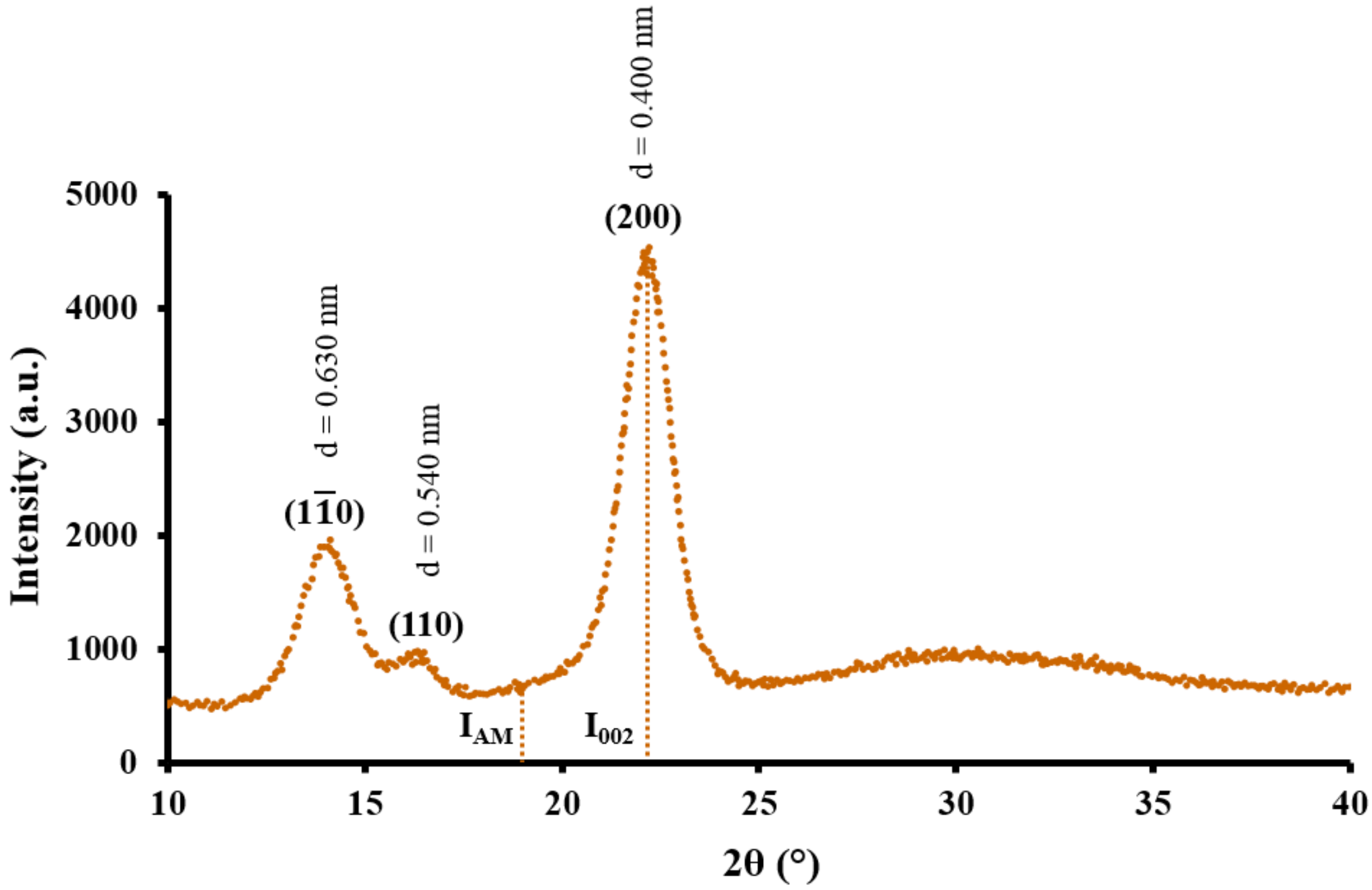

**Fig. S2** X-ray diffractogram of native BC membrane

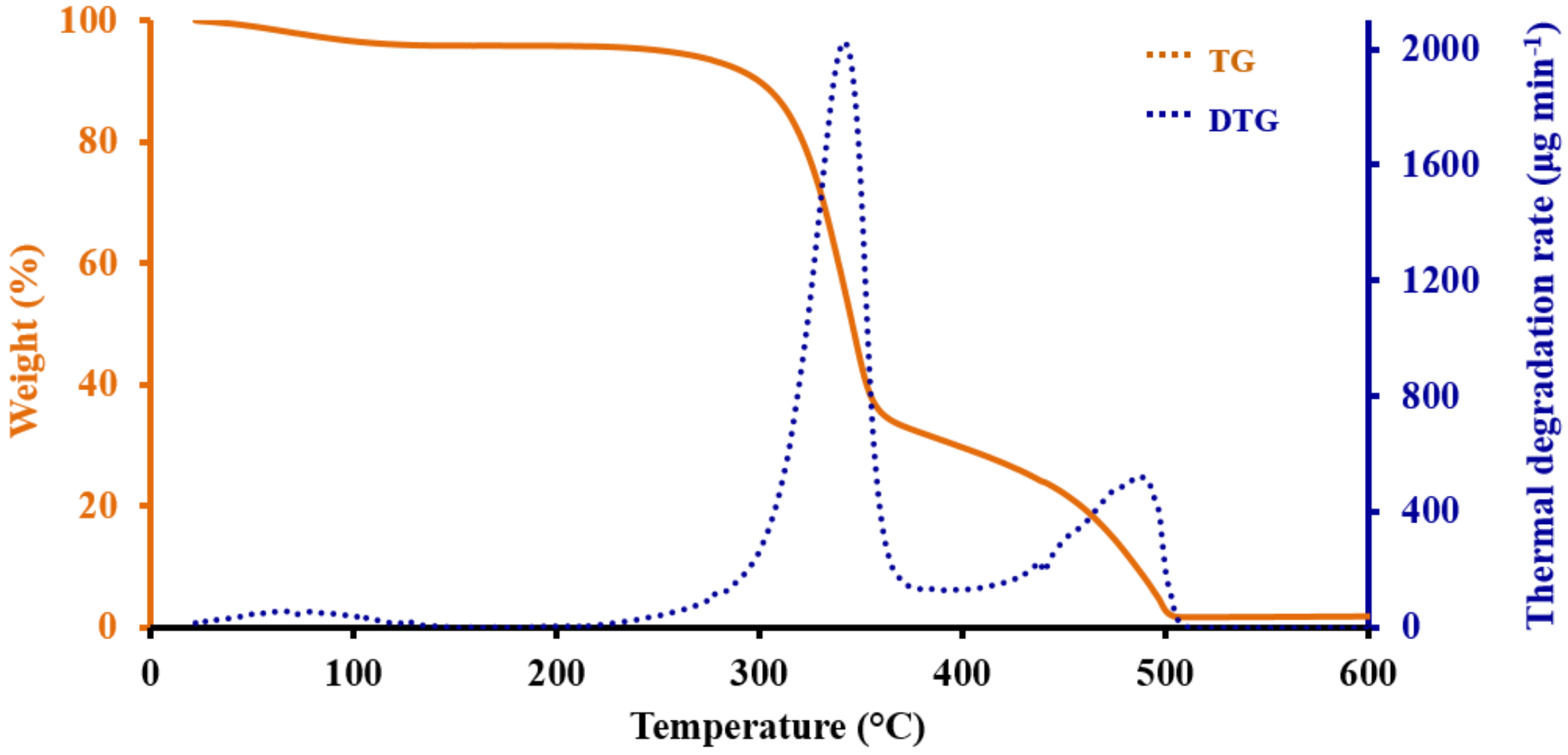

**Fig. S3** TG-DTG curve of native BC membrane

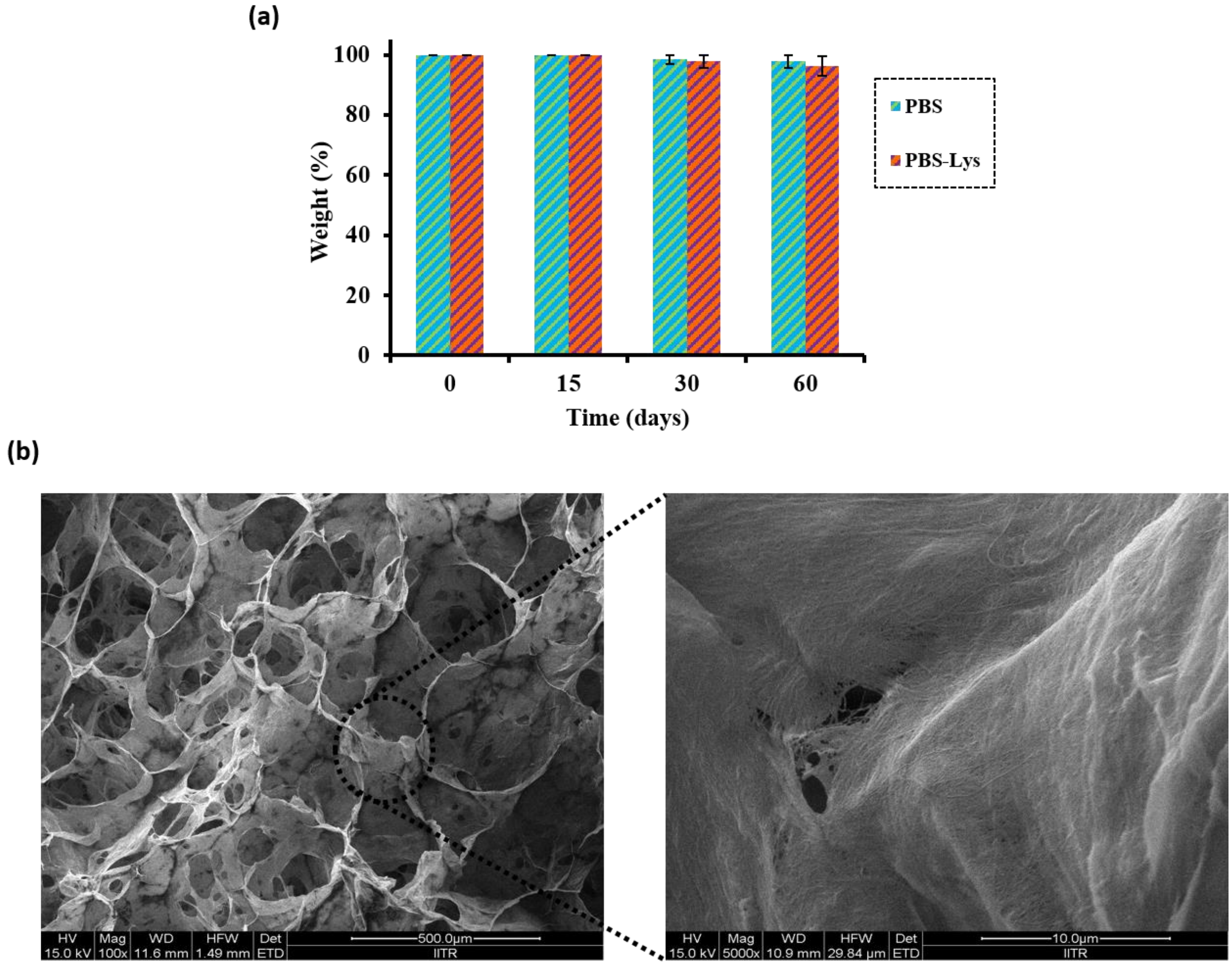

**Fig. S4** (a) Degradation behavior of mNBC scaffolds as a function of time; (b) microstructural morphology of mNBC scaffolds after two months of incubation in PBS-Lys at 37 °C. Values are represented as mean ± SD (n = 3) (PBS: Phosphate buffered saline; PBS-Lys: Phosphate buffered saline containing 0.2% lysozyme).

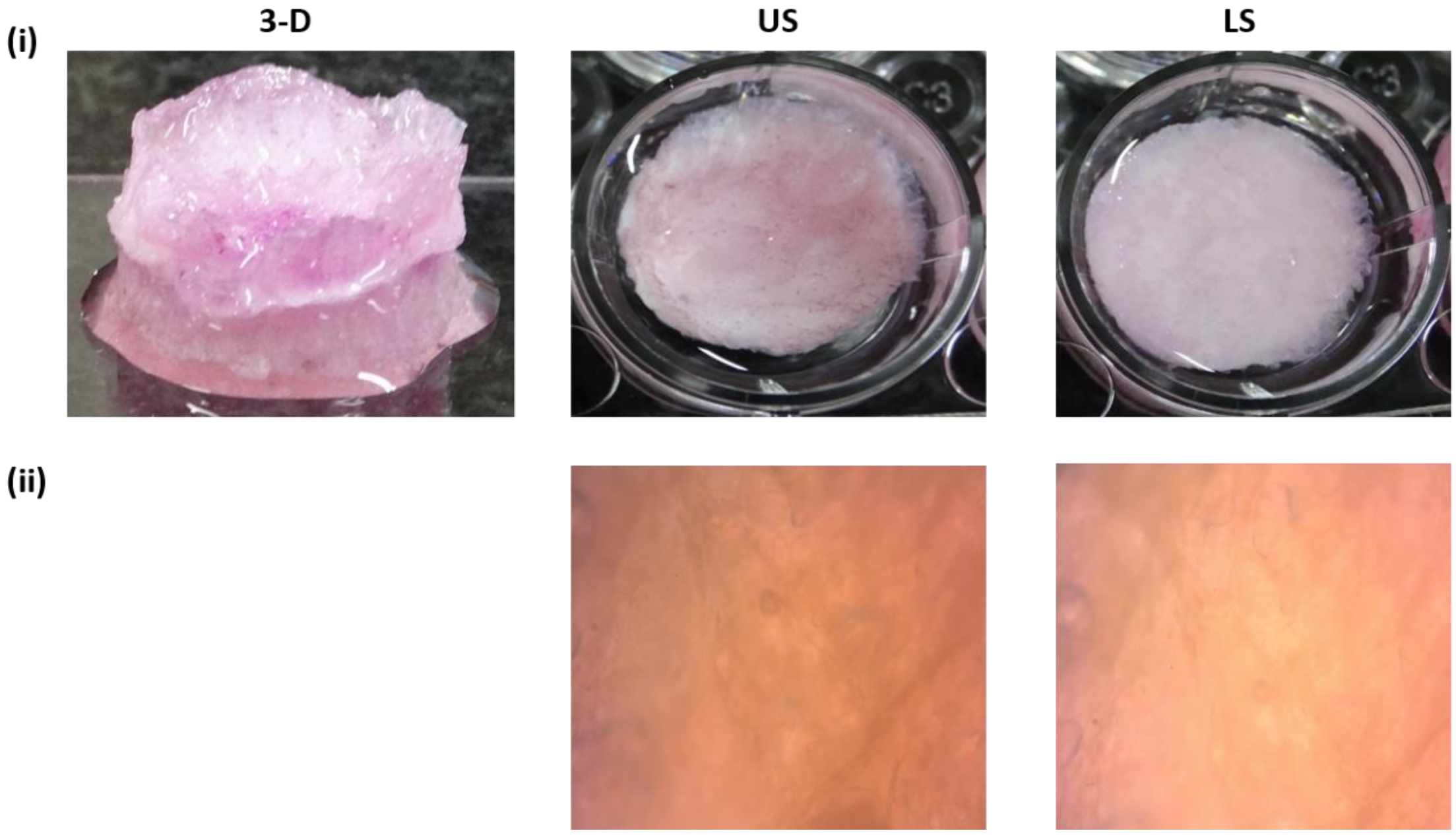

**Fig. S5** Alizarin red staining of mNBC scaffolds without cells. Panel (i) and (ii) shows the visual and microscopic images of the scaffolds, respectively. (3-D: 3-dimensional; US: upper side of the scaffold; LS: lower side of the scaffold)

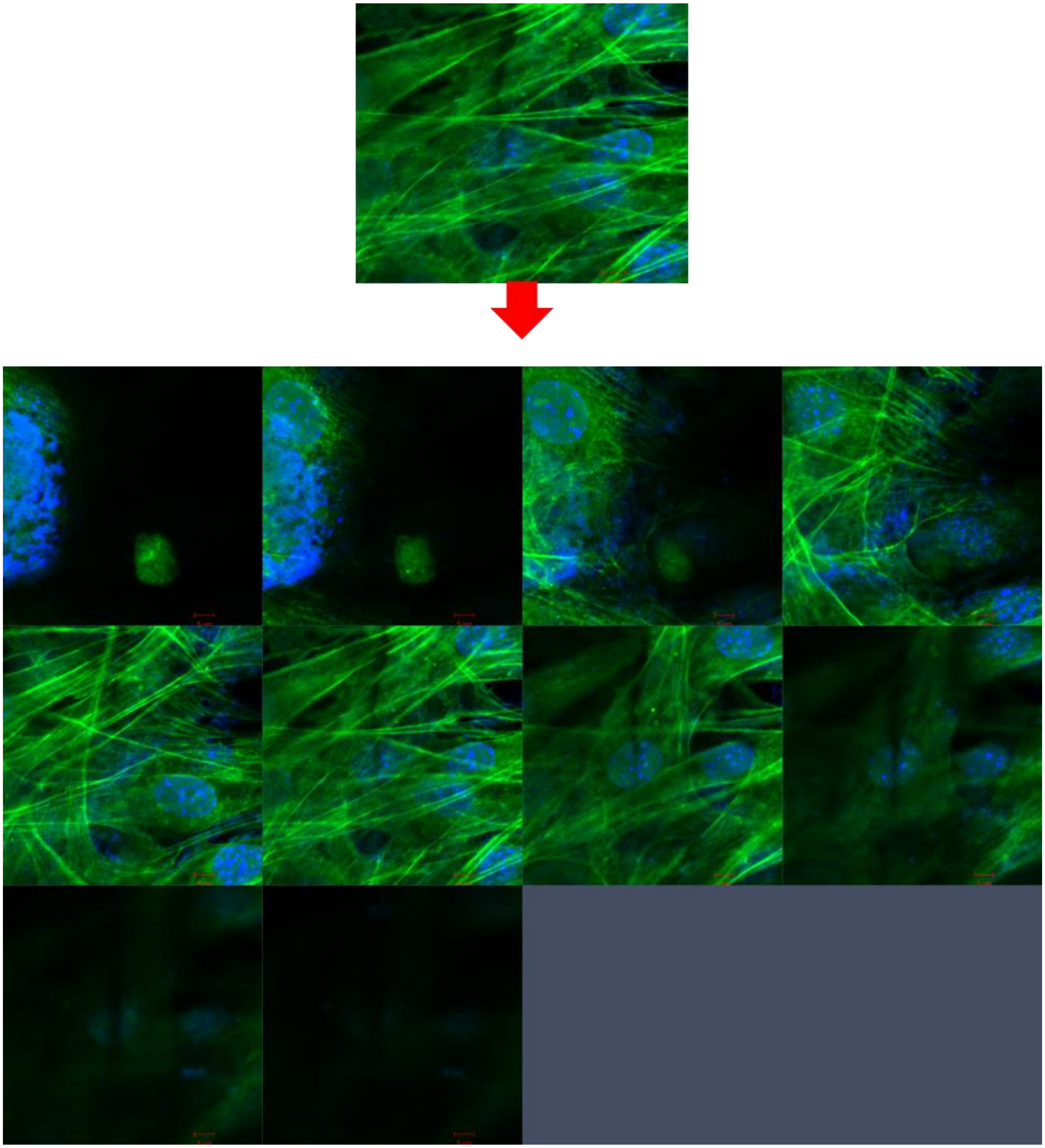

**Fig. S6** Confocal micrographs showing individual Z-stack images of phalloidin-FITC and DAPI stained C3H10T1/2 cells cultured on 3-D mNBC scaffolds after 21 days of culture in proliferation medium (optical slicing was done with 10 µm slice thickness up to 100 µm)

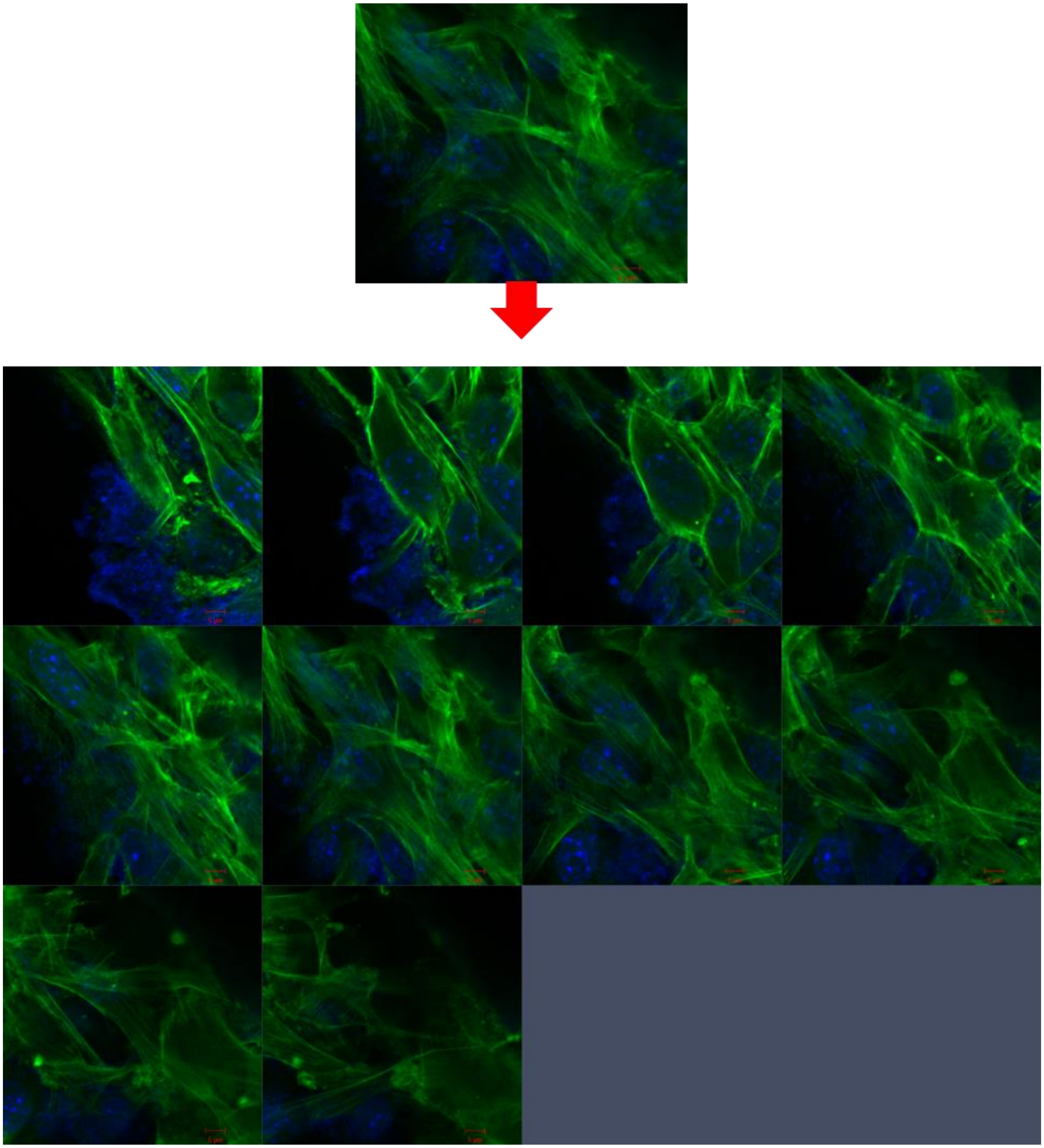

**Fig. S7** Confocal micrographs showing individual Z-stack images of phalloidin-FITC and DAPI stained C3H10T1/2 cells cultured on 3-D mNBC scaffolds after 21 days of culture in osteogenic medium (optical slicing was done with 10 µm slice thickness up to 100 µm)

## Graphical Abstract

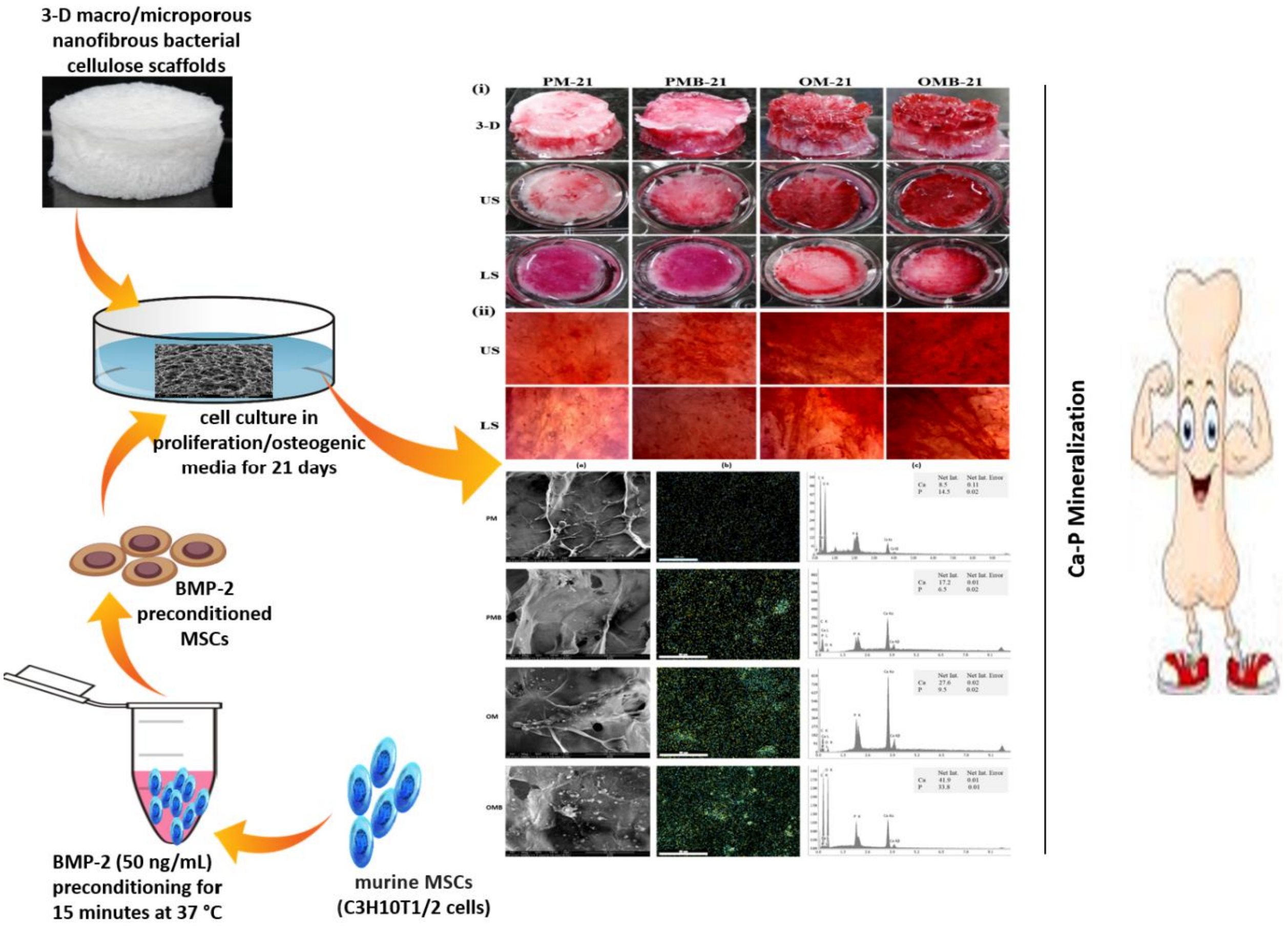